 \documentclass[final,5p,times,twocolumn]{elsarticle}

\usepackage{background}
\backgroundsetup{contents=PREPRINT}

\usepackage[T1]{fontenc}
\usepackage{amssymb}
\usepackage{amsthm}
\usepackage{amsmath}

\usepackage[utf8]{inputenc}
\DeclareUnicodeCharacter{00B3}{\textsuperscript{3}}
\usepackage{float}
\usepackage{mathtools}
\usepackage{braket}

\usepackage{isotope} 
\usepackage{bm} 
\usepackage{natbib}
\usepackage{booktabs} 
\usepackage{graphicx}

\usepackage{comment}

\usepackage{tabularx}
\usepackage{longtable}
\usepackage{multirow}
\usepackage{dcolumn} 
\usepackage{color} 
\usepackage[colorlinks,hyperindex]{hyperref} 

\usepackage[symbol]{footmisc}

\hypersetup
	{   colorlinks,%
		citecolor=blue,%
		linkcolor=blue,%
		urlcolor=blue,%
	}

\usepackage{lineno}

\journal{Nuclear Instruments and Methods A}

\begin{document}

\begin{frontmatter}




\title{Investigating timing properties of modern digitizers utilizing interpolating CFD algorithms and the application to digital fast-timing lifetime measurement}

\author[label1]{A. Harter\footnote[1]{Corresponding author}}
\ead{aharter@ikp.uni-koeln.de}

\affiliation[label1]{organization={Universitaet zu Koeln, Institut fuer Kernphysik},adressline={Zuelpicher Str. 77, 50937 Koeln, Germany}} 

\author[label1]{M. Weinert}
\author[label1]{L. Knafla}
\author[label1]{J.-M. R{e}gis}
\author[label1]{A. Esmaylzadeh}
\author[label1]{M. Ley}
\author[label1]{J. Jolie}

\begin{abstract}
The performance of two implementations of digital real-time interpolating constant fraction discriminator algorithms with respect to fast-timing lifetime measurements are investigated. The implementations integrated in two different digitizers were evaluated in terms of the effects of tuning parameters of the digital CFDs and the influence of different input amplitudes on the time resolution and time walk characteristics. Reference is made to the existing analog standard of fast-timing techniques. The study shows, that the timing performance of both modules is comparable to established fast-timing setups using analog constant fraction discriminators, but with the added benefit of digital processing. Both digitizer modules were found to be highly effective and user-friendly instruments for modern fast-timing requirements. 
\end{abstract}




\end{frontmatter}


\section{Introduction}
\label{sec:intruduction}
The fast-timing technique is a well known and reliable tool to determine lifetimes of nuclear excited states. Knowing the lifetime of an excited state of a nucleus can help in understanding underlying structures by extracting the reduced transition probability $B(\sigma L)$ between two excited states~\cite{casten2000nuclear,Krane1988}. Therefore, lifetime determination is a key ingredient to obtain structural information about excited nuclei. Throughout the evolution of nuclear physics, numerous techniques for determining lifetimes from seconds down to femtoseconds have been developed~\cite{Nolan.1979}. For lifetimes ranging from picoseconds to several nanoseconds, the fast-timing technique is a suitable method.\par
The fast-timing technique essentially utilizes the time difference between the populating and depopulating $\gamma$-rays of an intermediate state of interest to obtain information about its lifetime. The established fast-timing setup uses a complex circuit of analog constant-fraction discriminators (CFDs) and time-to-amplitude converters (TACs)~\cite{MACH.fasttiming,REGIS.yy-fasttiming}. However, using the multiplexed start-stop technique, the complexity of the electronics of an analog fast-timing setup  increases rapidly  with  the  number of detectors~\cite{REGIS.redyyTW.multiplexed,REGIS.GCD}.\par
In recent years, the emergence of digitally implemented real-time interpolating CFD algorithms integrated in fast-sampling digitizers has greatly reduced the complexity of electronic circuits, while providing timestamps in the picosecond range usable for fast-timing analyses. This study investigates the timing performance of the V1730 and V1751 digitizers manufactured by CAEN S.p.A. with sampling rates of 500~MHz and 1~GHz, respectively. The digitizers have on-board digitally implemented interpolating CFD algorithms to determine a timestamp of an incoming signal with a precision of around 2~ps or smaller~\cite{CAEN.1730,CAEN.1751,CAEN.compass}, enabling digital fast-timing with high precision.
The two digitizers with implemented interpolating CFD algorithms have been thoroughly investigated in terms of time resolution and amplitude-dependent time walk characteristics. A systematic investigation of the time walk behavior has been carried out. Additionally, the timing performance of both digitizers were compared to the characteristics of their analog counterparts.\par
The objective of this study is to comprehensively investigate the impact of different CFD tuning parameters and signal amplitudes on the time resolution and time walk characteristics and suggest a best set of parameters for the conditions present in this work.\par

\section{The fast-timing method}
\label{sec:ftmethod}
\subsection{General information about the fast-timing principle}
\label{ssec:ft-general}
The fast-timing procedure is based on measuring the time difference between two $\gamma$-rays of a $\gamma$-$\gamma$ cascade populating and depopulating an intermediate state of interest~\cite{MACH.fasttiming,REGIS.yy-fasttiming,REGIS.GCD, MACH.fasttiming2, MOSZYNSKI.fasttiming1,  REGIS.MSCD}.
The distribution of measured time-differences is described by the delayed time-difference distribution~\cite{BayCentroidPosition}. This delayed time-difference distribution is defined by a convolution of the prompt response function (PRF), which depends on the timing system, and an exponential decay~\cite{BayCentroidPosition}:
\begin{equation}
    D(t) = n\lambda \int_{-\infty}^t PRF(t'-C_P)e^{-\lambda(t'-C_P)}dt'+n_r, \qquad \lambda=\frac{1}{\tau},
\end{equation}
where $n_r$ is the random background level, $C_P$ is the centroid (first moment) of the PRF and $\tau$ the lifetime of the intermediate state. The PRF is the time difference distribution obtained for signals with a zero-time-difference. The centroid position and the full width at half maximum (FWHM) of the PRF, which is considered as the time resolution of a system, are dependent on detector properties, time pick-off technique (e.g. CFD, digital CFD or leading edge) and energies of detected $\gamma$-rays.\par
\begin{figure}[t]
\centering
    \includegraphics[width={250.00bp},height={150.00bp}]{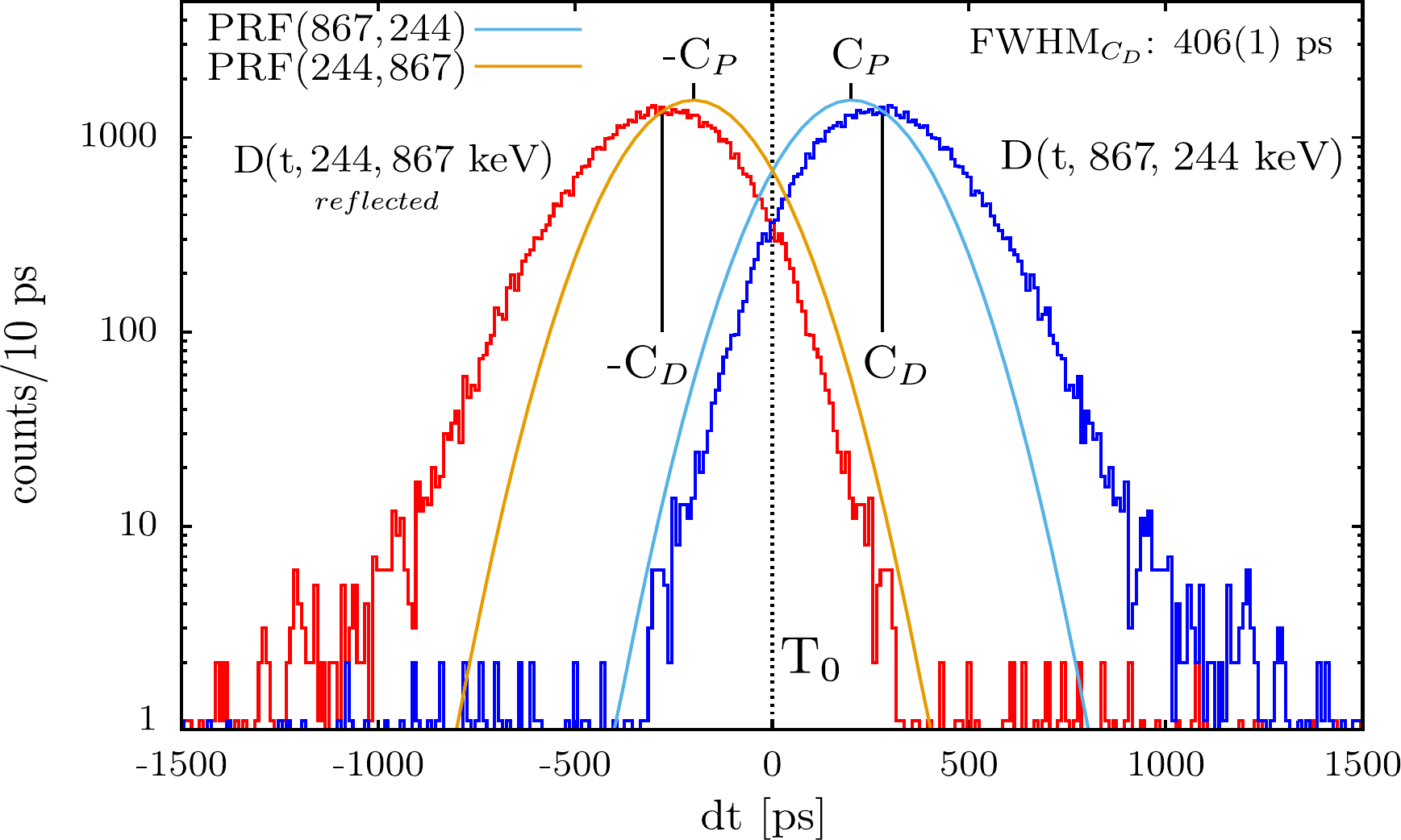}
    \caption{The identical but mirrored time difference distributions from the 867~-~244~keV cascade in \isotope[152]{Sm} with a binning of 10~ps/channel is shown. D(t,~867,~244~keV) (blue) is generated by gating in feeder - decay sequence. The reflected D(t,~244,~867~keV) (red) distribution is generated by gating in decay - feeder sequence. The intermediate state connecting both $\gamma$-rays has a lifetime of 80.9(11)~ps~\cite{Nucl.datasheet.152}. The time difference distributions resulting from the centroid shift method exhibit symmetry about the time reference point $T_0$ due to the inherent symmetry of the method.}
    \label{fig:244cascade-dists}
\end{figure}

The centroid-shift method is used to determine lifetimes of excited states with high precision, smaller than the time resolution of the corresponding experimental setup. In centroid-shift analysis, the lifetime is determined by the time shift of the centroids of the measured delayed time-difference distributions ($C_D$) relative to the centroid position of the energy-dependent PRF ($C_P$)~\cite{MACH.fasttiming}: $\tau = C_D - C_P$. This situation is illustrated in Fig.~\ref{fig:244cascade-dists}, where the centroid positions considered are indicated. Further details can be found in Refs.~\cite{REGIS.yy-fasttiming,REGIS.GCD}. \par
The centroid position of the PRF is affected by signal amplitude depended time walk (TW).
If the TW is calibrated for the experimental setup, the lifetime of an intermediate state can be derived according to:
\begin{equation}
    \tau = C_D - T_0-TW(E_1,E_2).
    \label{eq:tau}
\end{equation}
The constant term T$_0$ is given by $T_0 = TW(E_1 = E_2)$, which corresponds to the zero point of the timing system.

\subsection{From analog to digital timing}
\label{ssec:two-diff-appr}
The technical state-of-the-art for lifetime measurements using large fast-timing arrays in terms of time resolution, time walk behavior is the use of analog setups using multiplexed start and stop signals~\cite{REGIS.redyyTW.multiplexed}. This established analog fast-timing setup is constructed using two or more fast-timing detectors, typically LaBr$_3$(Ce) or CeBr$_3$. These scintillators are optically coupled to photomultiplier tubes (PMTs)~\cite{POLYAKOV.PMT, REGIS.PMT.Phd} with two outputs each. The analog setup utilizes two branches, a time branch and an energy branch. The energy branch directly digitizes the energies of the incoming $\gamma$-rays. The time branch uses a circuit of analog Fan-in-Fan-out modules, delay loops, CFDs and TACs, to determine the time difference $dt$ between two signals~\cite{Nolan.1979,MACH.fasttiming}. Therefore, the information of a single fast-timing event is condensed to
$$(\textrm{E}_1, \textrm{E}_2, dt),$$
where the energies E$_1$ and E$_2$ are obtained from the energy branches and the time difference $dt$ from the TAC in the time branch. This technique is described in detail in Refs.~\cite{MACH.fasttiming,REGIS.yy-fasttiming,REGIS.GCD,REGIS.MSCD}. Utilizing modern TACs and CFDs this kind of analog setup is able to measure short lifetimes down to a few picoseconds.\par
In various measurements, tests and quantifications, the analog CFD model ORTEC 935~\cite{ORTEC.935} in combination with the TAC model ORTEC 566~\cite{ORTEC.566} has proven high fast-timing performance~\cite{REGIS.PMT.Phd,Esmayl.fasttiming2,Harter.fasttiming,karayon.fasttiming,knafla.fasttiming,AEsmayl.fasttiming,vkarayon.fasttiming}. This combination of analog electronics represents the state-of-the-art of fast-timing technology and serves as the standard for comparison in this work.\par
In the following, the term "analog setup" refers to a configuration in which time difference information is obtained from TACs, while the term "digital setup" refers to a fully digital configuration in which analog electronic modules are not present and the direct time information of signals is measured by the digital CFDs.\par
Recent fast-sampling digitizers utilizing digitally implemented interpolating CFD algorithms (described in Sec.~\ref{sec:cfd}) are capable of determining sub-sample-period precision timestamps. The high precision of digital timestamp determination in the low picosecond regime and the omission of the external analog modules enables an extensive simplification of fast-timing setups. In a fully digital experimental setup, the information of a single fast-timing event is now condensed to 
\begin{equation}
    \bigl[(\textrm{E}_1, t_1), (\textrm{E}_2, t_2)\bigr],
    \label{eq:event}
\end{equation}
containing the full measured information. The time difference is taken in the offline analysis in both possible directions~\cite{REGIS.symmetricCube}:
\begin{equation}
    dt_{12}(\textrm{E}_1, \textrm{E}_2) = t_2 - t_1 \textrm{ ~and~ } dt_{21}(\textrm{E}_2, \textrm{E}_1) = t_1 - t_2.
    \label{eq:dt}
\end{equation}
Conceptual, the analysis of the digital fast-timing data is identical to centroid shift analysis of $\gamma$-$\gamma$ coincidence data and is discussed in detail in Refs.~\cite{MACH.fasttiming,MACH.fasttiming2,REGIS.symmetricCube}. This symmetrization approach is appropriate when nearly equal timing conditions, e.g. similar time resolutions of detectors, mainly related to the crystal size, and time walk, mainly related to the CFD device, are given for each channel.\par
By applying energy conditions as gates to the collection of all stored events $\bigl[(\textrm{E}_1, t_1), (\textrm{E}_2, t_2)\bigr]$ and calculating $dt$ according to Eq.~\ref{eq:dt}, a time difference distribution is generated. In Fig.~\ref{fig:244cascade-dists}, two distributions are shown, both containing the full experimental data. The distributions are identical but mirrored with respect to T$_0$. Only the gate sequence, either (E$_{\textrm{feeder}}$,~E$_{\textrm{decay}}$) or (E$_{\textrm{decay}}$,~E$_{\textrm{feeder}}$), determines, which distribution is generated. In Fig.~\ref{fig:244cascade-dists}, this process is illustrated for the 876-244~keV cascade in \isotope[152]{Sm} with a lifetime of the intermediate state of 80.9(11)~ps~\cite{Nucl.datasheet.152}, populated by the electron capture decay from  \isotope[152]{Eu}. The symmetry of the distributions under exchange of E$_1$ and E$_2$ can be used for precise determination of T$_0$ of the system. From the distributions in Fig.~\ref{fig:244cascade-dists}, e.g. D(t,~E$_{\textrm{feeder}}$,~E$_{\textrm{decay}}$) the lifetime can be extracted using the centroid shift method by applying Eq.~\ref{eq:tau} with known time walk characteristics.\par

\section{Digital constant fraction discrimination and timestamp interpolation}
\label{sec:cfd}
\begin{figure}[t]
\centering
    \includegraphics[width={250.00bp},height={150.00bp}]{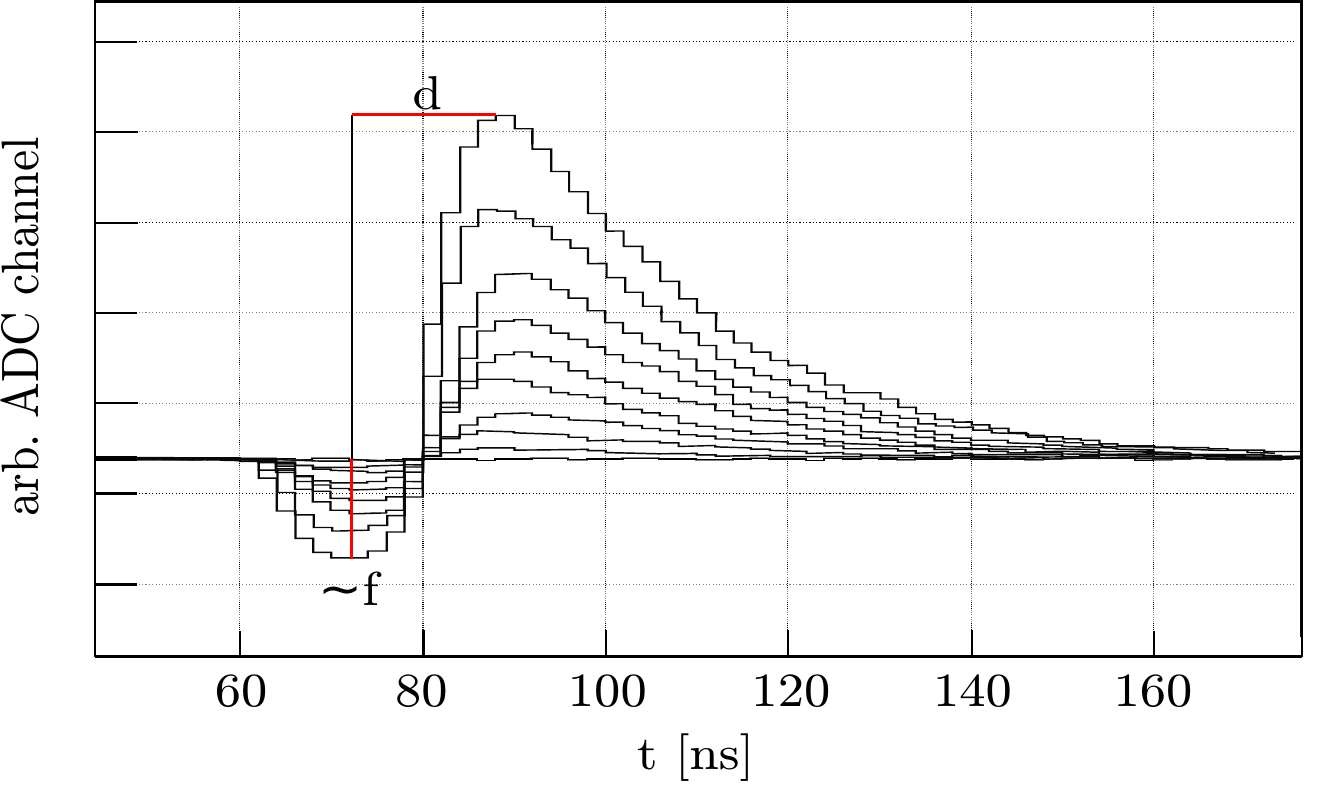}
    \caption{Several digitally CFD shaped signals recorded by a CAEN V1730 digitizer module. The input signal was a negative anode signal originating from a LaBr detector in combination with a PMT. In this exemplary case, $d$ marks the delay time and $\sim f$ an approximation of the fraction value used in the shaping procedure.}
    \label{fig:stacked-cfd}
\end{figure}

\begin{figure}[t]
\centering
    \includegraphics[width={250.00bp},height={120.00bp}]{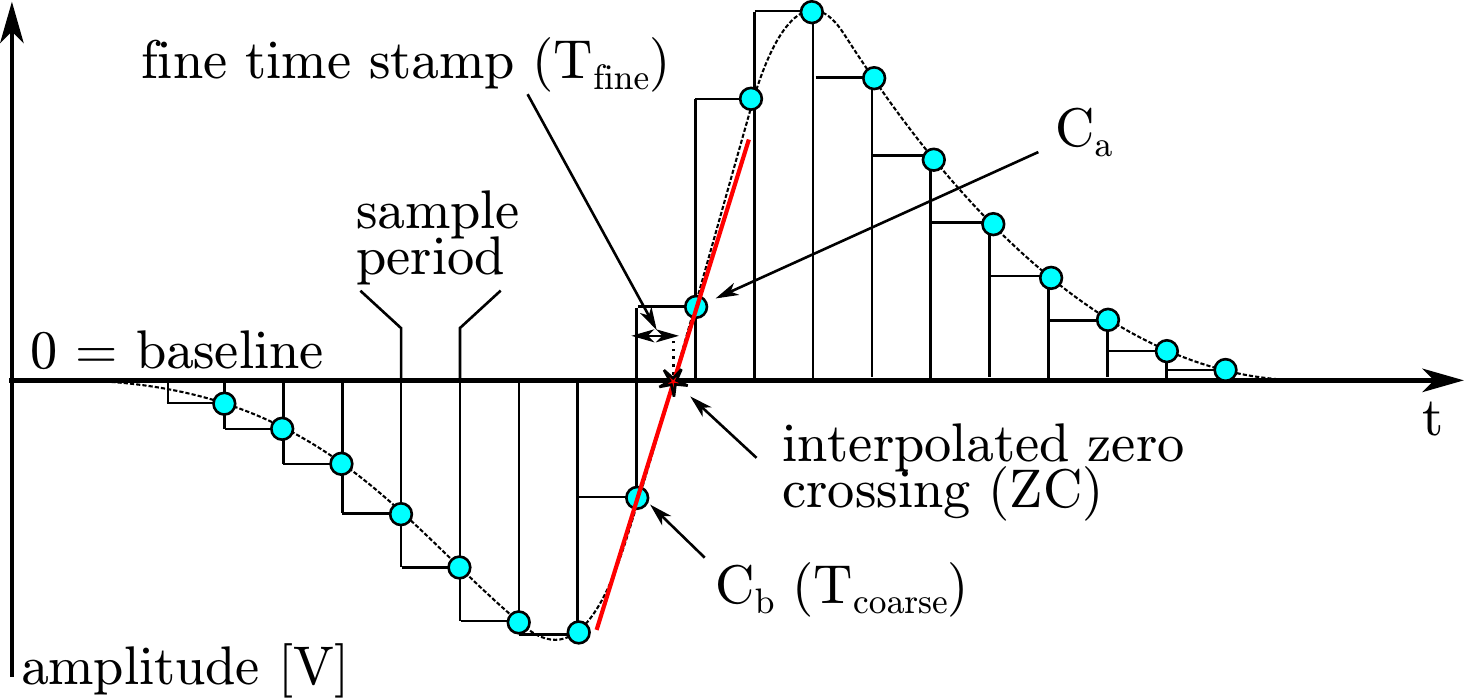}
    \caption{A sampled CFD signal, where the interpolated polynomial of first order is indicated by a red line and the zero crossover marked by a star. The point in time of the zero crossover represents the timestamp of the original input signal with a precision smaller than the sample period.}
    \label{fig:interp}
\end{figure}

A fast-sampling analog-to-digital converter (ADC) is capable of accurately discretizing continuous input signals that have durations within the nanosecond range. This enables a digital CFD signal shaping procedure to achieve a zero crossover at the point in time the original signal reaches a constant fraction of its amplitude. The timestamp of an incoming signal is then identified as the zero crossover of a CFD shaped signal. The digital shaping procedure of the implemented CFD algorithm is analogous to the shaping procedure of an analog CFD. Specifically, the captured signal is duplicated, one of the duplicates is inverted and delayed by a specified delay time $d$, while the other one is attenuated by a defined fraction $f$~\cite{Paulus.CFD}, see Fig.~\ref{fig:stacked-cfd}.
The presented algorithm superimposes two processed signals to create a bipolar signal, which contains the required zero crossover to calculate a constant fraction timestamp according to~\cite{FALLULABRUYERE.CFDformel}:
\begin{equation}
    C_1(t) = S(t)\cdot f-S(t+d)
\label{eq:cfd1}
\end{equation}
    or
\begin{equation}
    C_2(t) = S(t+d)-S(t)\cdot f,
    \label{eq:cfd2}
\end{equation}
where $C_n(t)$ is the CFD sample and $S(t)$ the original sample at time $t$. A sample is a measurement of the amplitude of an analog signal taken at discrete time intervals known as the sampling period. The ADC then converts each sample into a digital representation, which results in the discretization of the continuous signal pulse. The two distinct shaping sequences, as presented by Eq.~\ref{eq:cfd1} and Eq.~\ref{eq:cfd2}, represent two different implementations found in the digitizer modules investigated in this study. Figure~\ref{fig:stacked-cfd} shows a series of digitally shaped CFD signals with different amplitudes originating from the negative anode of a LaBr scintillator coupled to a PMT. These signals were recorded with a CAEN model V1730 digitizer.\par
The timestamp derived from the zero crossover would have a precision equal to the sample period, e.g. 2~ns for a 500~MS/s sample frequency. But to achieve timestamps with higher precision than the sample period, interpolations of low-degree polynomials between adjacent samples surrounding the zero crossover of the CFD signals are utilized. 
Recent developments in field-programmable gate array (FPGA) technology allow the real-time interpolation of such polynomials already on the digitizer and online sub-sample-period timestamp calculation. Recently, various interpolation algorithms including linear and cubic spline interpolations heve been discussed in~\cite{LIPSCHUTZ.splitsignal,MODAMIO.cfd,SANCHEZ.cfd,warburton.timingAlg}.\par
The digital CFDs of the modules under investigation in this study interpolate a first-order polynomial between the last CFD sample before the zero crossover $C_b$ and the first CFD sample after the zero crossover $C_a$. As illustrated in Fig.~\ref{fig:interp}, the timestamp of the interpolated zero crossover (ZC) is expressed by the sum of the timestamp of $C_b$ ($T_{coarse}$, with sample period precision) and the fine timestamp T$_{\textrm{fine}}$:
$$
\textrm{ZC} = T_{\textrm{coarse}} + T_{\textrm{fine}}.
$$ 
T$_{\textrm{fine}}$ is calculated using the linear interpolation between $C_b$ and $C_a$ according to the linear equation:
\begin{equation}
    T_{\textrm{fine}} = \frac{-C_b}{C_a-C_b} \cdot T_{\textrm{sample}}.
\end{equation} 
$T_{\textrm{sample}}$ is the sample period determined by the sample rate of the data acquisition, e.g. 2~ns for a sample rate of 500~MS/s.\par

In analog CFD modules, the time resolution strongly depends on the amplitude of the input signal and the amplitude resolving capability of the electronic components used in the CFD circuit~\cite{Paulus.CFD,REGIS.CFD}. This effect depends on the signal-to-noise ratio (SNR)~\cite{Paulus.CFD}, which is defined as the ratio of the signal amplitude to any electronic noise present in the signal. The SNR induces an additional time uncertainty which is significant at lower amplitudes since the slope at the zero crossover decreases with decreasing amplitudes. A similar effect is expected for a digitally shaped CFD signal and applied timestamp interpolation. 
\begin{figure}[t]
\centering
    \includegraphics[width={250.00bp},height={150.00bp}]{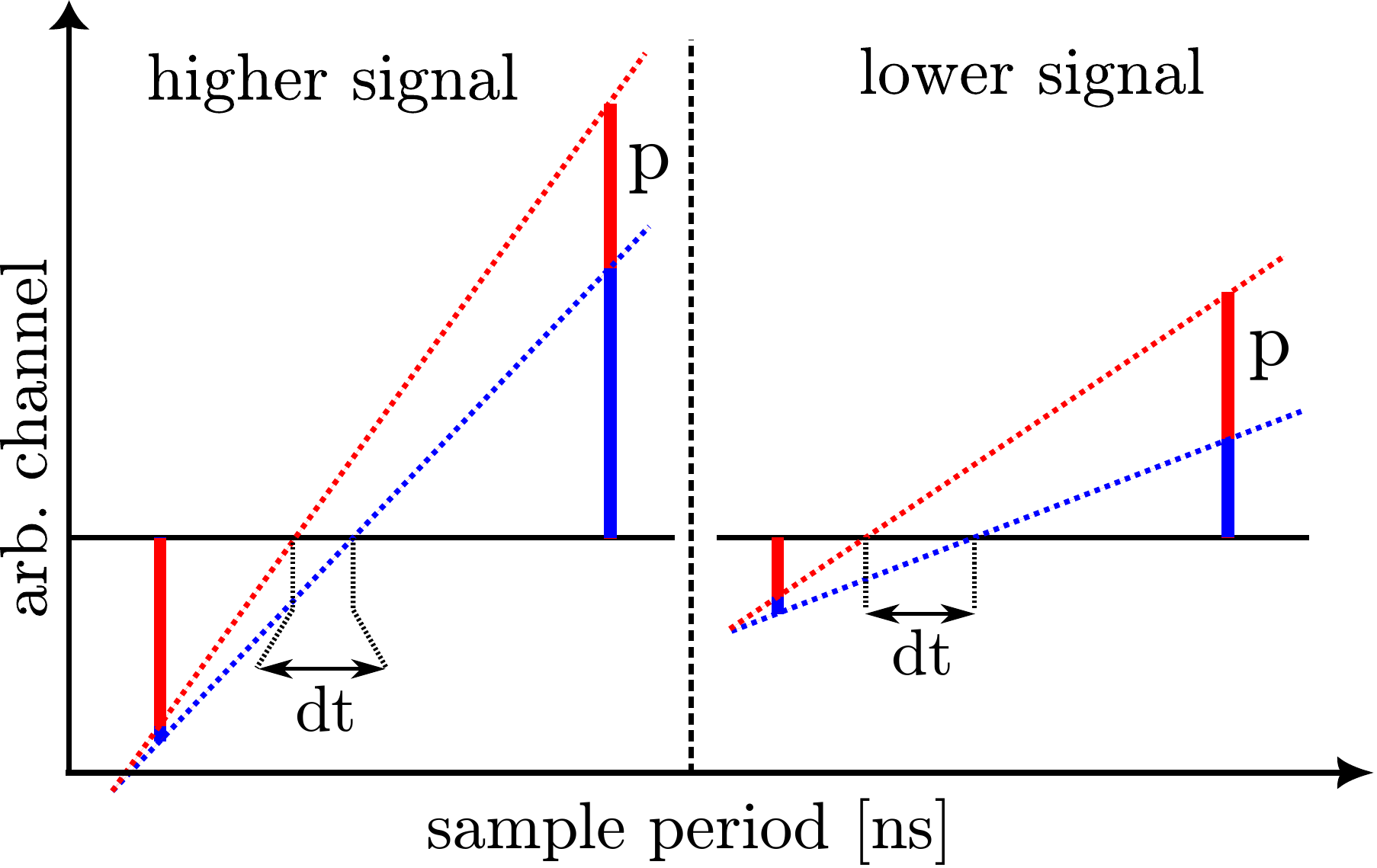}
    \caption{Illustration of two different signal amplitudes affected by the disturbance $p$ and effect on the timestamp interpolation. The disturbance $p$ of the sample amplitudes, considered as noise, is the same on the larger (left) and on the lower (right) signal. The disturbance was chosen to be relatively large in order to demonstrate the effect. The interpolated timestamp of the higher signal differs in a smaller range $dt$ than that of the smaller signal. As a result, small variations of small signals lead to larger uncertainties in the interpolation of the timestamp compared to higher signals, which is called time jitter.}
    \label{fig:time-resol}
\end{figure}
As depicted in Fig.~\ref{fig:time-resol}, an artificial disturbance $p$ added to two different samples affects the interpolated timestamp depending on the amplitude. The smaller the amplitude, the larger the deviations of the timestamps $dt$. This causes the time resolution of digital CFDs to degrade for smaller signal amplitudes. This effect is also referred as time jitter~\cite{Paulus.CFD}.\par 
Digital CFDs, as well as analog CFDs, are affected by an energy dependent time walk (see Sec.~\ref{ssec:ft-general}). The time walk of an analog CFD occurs due to the slope of the bipolar CFD signal at the point of zero crossover~\cite{Paulus.CFD}. The slope at this point decreases rapidly with decreasing amplitude for amplitudes below around 400~mV~\cite{REGIS.redyyTW.multiplexed}. Above approximately 400~mV, the slope remains nearly constant~\cite{REGIS.redyyTW.multiplexed}. The non-linear time walk at low amplitudes is caused by the charge sensitivity of the electronics within the CFD module, while the nearly linear time walk at higher amplitudes is related to the shaping delay time~\cite{Paulus.CFD,REGIS.CFD}. It is reasonable to expect similar behavior of slope-dependent time walk in digital interpolating CFDs.\par

\section{The sorting procedure of digital fast-timing data}
\label{sec:code}
\begin{figure}[t]
\centering
    \includegraphics[width={250.00bp},height={182.00bp}]{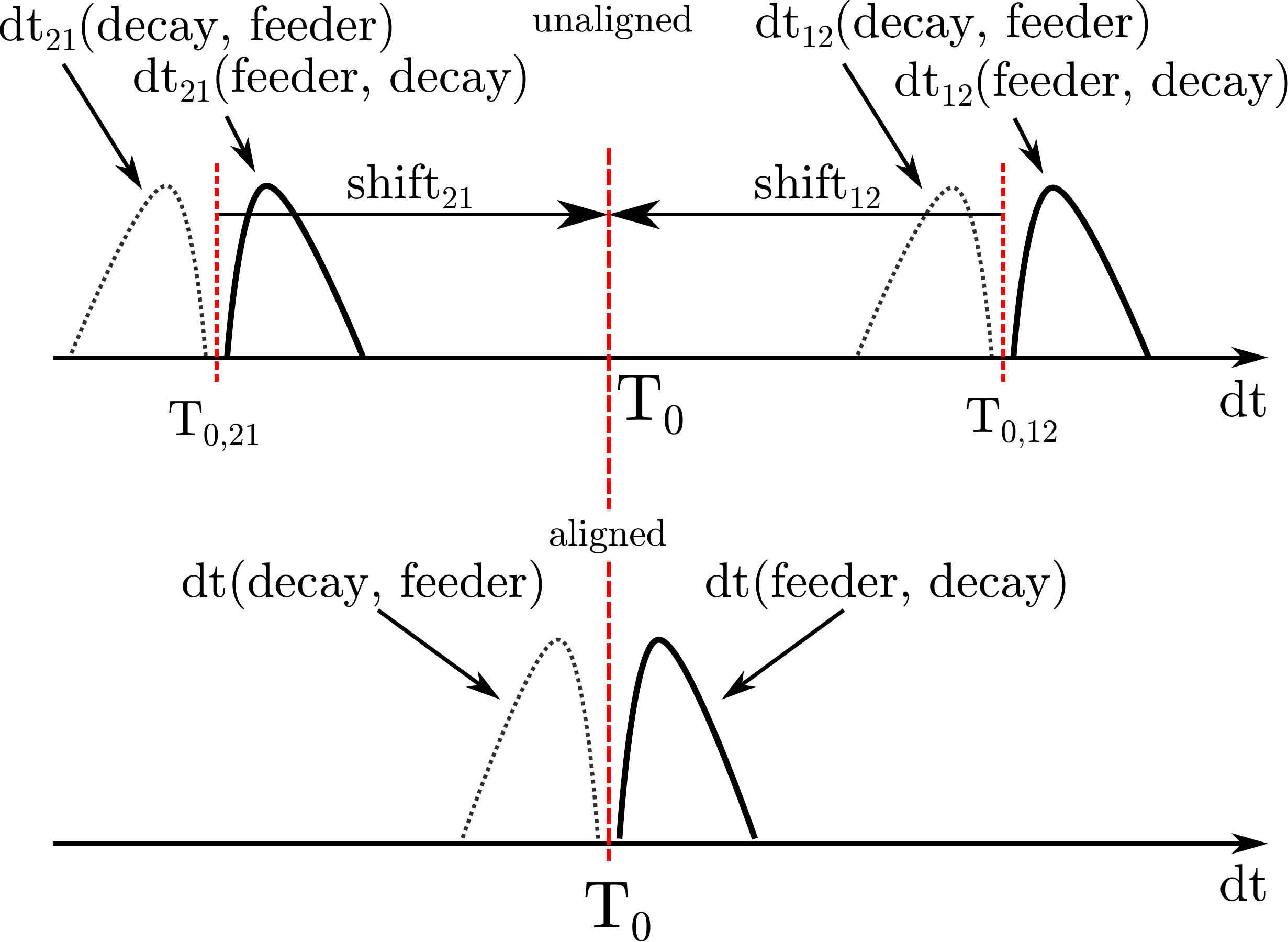}
    \caption{Digital timestamp shift for the detector combination $ij=12$. After the application of the shift constants, all time difference distributions are aligned. This is done for each detector pair $ij$. The dashed time difference distribution is the anti-delayed distribution, obtained by exchange of the gates. In the lower, aligned plot, both distributions are identical but mirrored with respect to T$_0$ and each contains the full experimental statistics.} 
    \label{fig:dtshift}
\end{figure}
In digital fast-timing, the time difference of two time correlated $\gamma$-rays is extracted from the precise timestamp information of each individual detector hit, which is done during the offline analysis. All detector hits are sorted by timestamp after the determination of potential runtime differences between individual channels outside the data acquisition system, e.g., due to different cable lengths.

For symmetric data of a digital setup, the time difference distributions for a single detector combination, calculated according to Eq.~\ref{eq:dt}, have to be aligned by applying a set of $shift_{ij}$ constants to each distribution before superposition of the data from all detector combinations.
The procedure is illustrated in Fig~\ref{fig:dtshift} and is representative of the alignment of all time spectra $dt_{ij}$, ($i$, $j \in N~:~ i \ne j$) of an array with N detectors~\cite{REGIS.symmetricCube}. To determine the $shift_{ij}$ constants precisely, the time difference distributions of a cascade with a short, nearly prompt lifetime, like 779-344~keV is utilized. The energy-independent shift constants are obtained by half the distance of $dt_{12}$ and $dt_{21}$, where T$_{0,ij}$ and T$_{0,ji}$ are symmetric around T$_0$, according to Fig~\ref{fig:dtshift}:
$$shift_{ij} + dt_{ij} = shift_{ij} + t_i - t_j = dt$$
and
$$ shift_{ji} + dt_{ji} = shift_{ji} + t_j - t_i = dt.$$
Once the time difference distributions have been aligned, the relationship $dt_{ij} = dt_{ji} = dt$ is established. Figure~\ref{fig:dtshift} shows that for a given detector combination $ij$ and the following relationship holds: $shift_{ij} = -shift_{ji}$. Different detector combinations $ij$ exhibit different shifts relative to the reference time $T_0$.


All detector hits are sorted by applying a time coincidence window and multiplicity conditions and stored as events. A minimal digital fast-timing event has multiplicity two and consists of two energy values and corresponding timestamps according to Eq.~\ref{eq:event}.\par
After identifying the events of interest by using the energy conditions, the desired time differences are calculated using Eq.~\ref{eq:dt}. Note, for consistency we opted to always calculate the time difference according to
$$
dt = t_{\text{E}_{\text{second gate matching}}} - t_{\text{E}_{\text{first gate matching}}},
$$
where suffix "first gate matching" corresponds to the first energy condition and "second gate matching" to the second energy condition applied to the data. This method ensures that the time difference is calculated in a consistent manner resulting in the desired distributions, as illustrated in Fig.~\ref{fig:244cascade-dists}.\par
\begin{figure}[t]
\centering
    \includegraphics[width={250.00bp},height={260.00bp}]{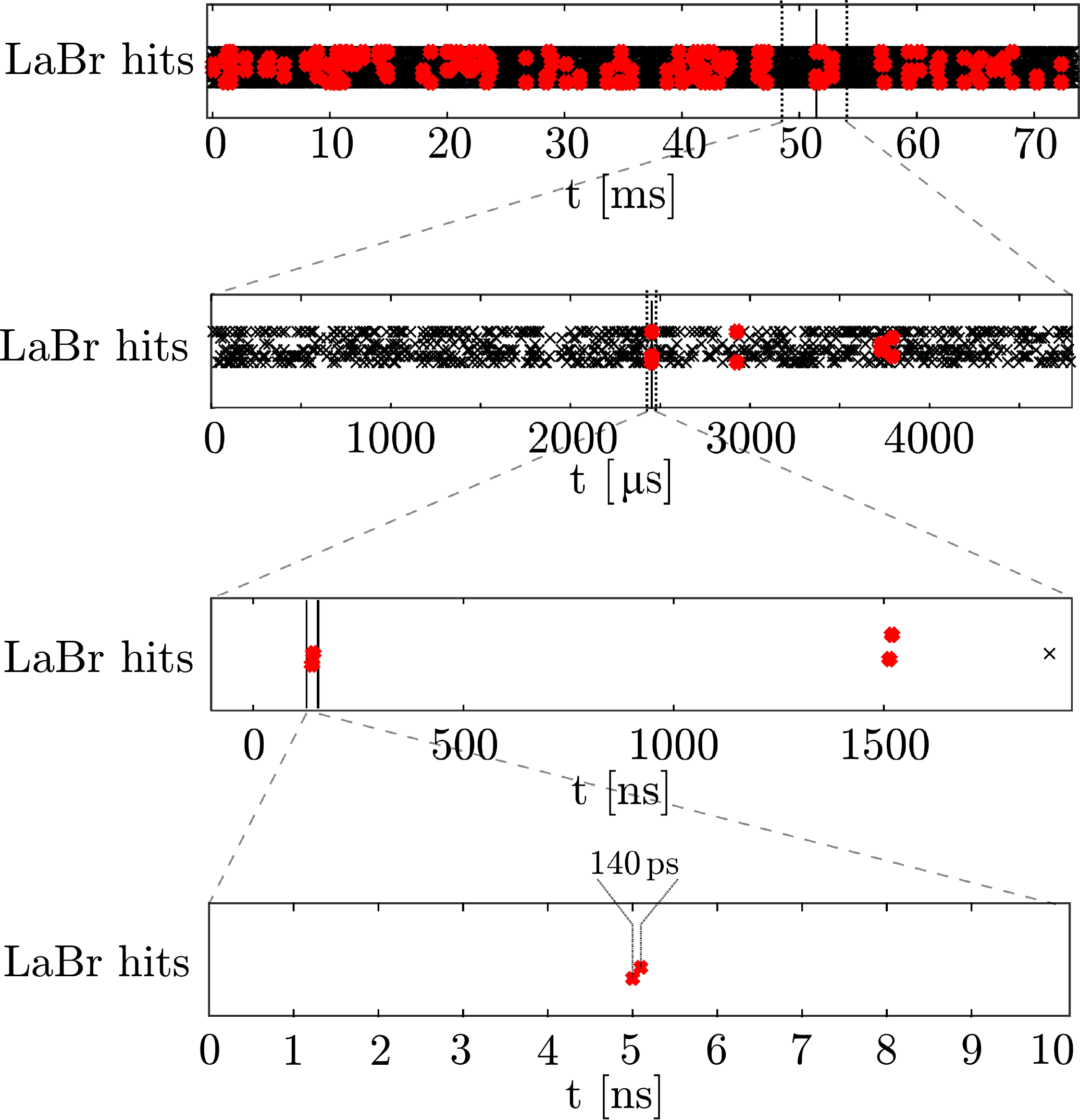}
    \caption{Detector hit density for different time ranges between 70~ms to 10~ns of a fast-timing array with 8 LaBr detectors. The data was measured during an in-beam experiment with detection rates of 15-20~kHz per detector. The data is sorted with a coincidence window of 10~ns and a multiplicity of exactly two LaBr hits was required. The red crosses represent detector hits that satisfy these conditions. The two LaBr hits in the bottom plot have a time difference of only 140~ps. The figure demonstrates the accuracy of the digital timestamp determination and motivates the use of short coincidence windows of around 10~ns for digital LaBr-LaBr fast-timing. Note: the time axis were reset to zero in each plot to provide reasonable time ranges in the individual plots.}
    \label{fig:zoom}
\end{figure}
Figure~\ref{fig:zoom} depicts the event density in an example time range of 70~ms with different zoom levels centered on one specific event of multiplicity~=~2. This data was recorded during an in-beam experiment using a fast-timing array equipped with 8 LaBr scintillation detectors and detection rates of 15-20~kHz per detector. The red crosses indicate the detector hits that satisfy the conditions of a coincidence window of 10~ns and a multiplicity of exactly 2 LaBr hits. The black crosses represent all other detector hits uncorrelated in time. The figure demonstrates the accuracy of recent digital CFDs for determining timestamps. It highlights the importance of an accurate timestamp shift correction, as a coincidence window in the order of a few nanoseconds can be sufficient to measure short lifetimes in the picosecond range, as long as only LaBr hits or those of comparable fast detectors are used.\par


\section{Experimental details}
\label{sec:exp}
This study aims to investigate the fast-timing capabilities of modern digitizers, namely CAEN modules V1730~\cite{CAEN.1730} and V1751~\cite{CAEN.1751}, implementing digital interpolating CFD algorithms for picosecond precise timestamp determination of detector hits. The important hardware properties of these digitizers concerning digital $\gamma$-$\gamma$ fast-timing are listed in Tab.~\ref{tab:digitizers}. The goal of this part of the study is to optimize the time resolution of the digital CFDs in combination with the established LaBr detectors while minimizing the energy-dependent time walk in the $\gamma$-ray energy range of a \isotope[152]{Eu} time walk calibration standard. An optimization process was conducted to identify the optimal combination of CFD algorithm parameters (see Sec.~\ref{ssec:approaches}) providing the best balance between time resolution and minimal time walk. 
\begin{table}[t]
    \caption{Resolution of the ADCs, internal sample rate and input dynamic range for the CAEN V1730 and V1751 digitizers.}
    \centering
    \begin{tabular}{cccc}
     digitizer  & ADC res. & sample rate & input range \\
     \hline
     V1730    & 14~bit & 500~MS/s & 0.5/2~Vpp \\
     V1751 & 10~bit & 1~GS/s & 1~Vpp
    \end{tabular}
    \label{tab:digitizers}
\end{table}

\subsection{Digital fast-timing setup}
\label{ssec:testsetup}
\begin{figure}[t]
\centering
    \includegraphics[width={206.00bp},height={215.00bp}]{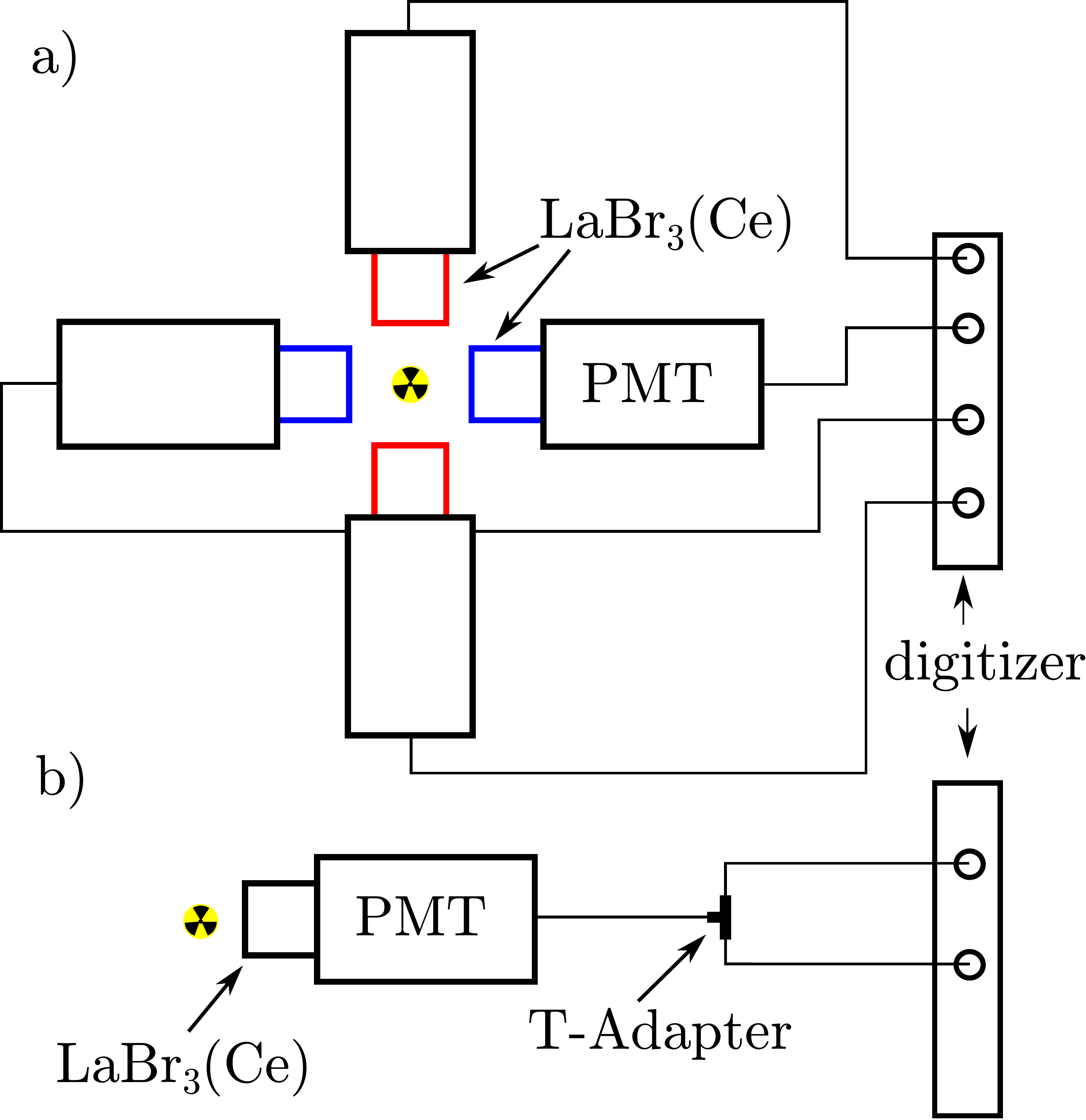}
    \caption{(a) Fast-timing setup with four LaBr$_3$(Ce) scintillators in a $90^{\circ}$ geometry with a source-to-detector distance of 2.5~cm. Only $\gamma$ - $\gamma$ coincidences between opposing detectors were used, which is indicated by the colors of the scintillators. This geometry grants good exposure of the scintillators and low inter-detector Compton scattering. (b) Split signal setup using one LaBr detector and a T-adapter. The two signals are fed into two different digitizer channels.}
    \label{fig:setup}
\end{figure}

In order to evaluate the timing performance of the implemented interpolating CFD algorithms of the V1730 and the V1751 digitizers, a fast-timing setup was constructed as shown in Fig.~\ref{fig:setup}a. The setup consists of four 1.5'' x 1.5'' LaBr$_3$(Ce) scintillators optically coupled to Hamamatsu R13435 photomultiplier tubes~\cite{Hamamatsu.pmt}. Each detector was connected to the digitizer with a single cable from the negative anode output of the PMT. A $90^\circ$ geometry with a distance of approximately 5~cm between the detectors was applied, with only opposite detectors being used for timing coincidences. This geometry allows for a good balance between reduced inter-detector Compton scattering and sufficient statistics and efficiency~\cite{REGIS.yy-fasttiming,REGIS.GCD}. Both digitizer modules support the same CFD shaping parameters in terms of timing capabilities: the "CFD delay", which corresponds to the delay time $d$, and the "CFD fraction", which corresponds to the constant fraction $f$, both described in Sec.~\ref{sec:cfd}. The CFD delay is adjustable in units of the sample period. The CFD fraction is selectable between four values (25\%, 50\%, 75\%, and 100\%) for the applied firmware of these particular modules.\par
An \isotope[152]{Eu} $\gamma$-ray source was used in the experiments, which offers twelve nearly Compton background-free $\gamma$-ray cascades with known lifetimes of intermediate states in the energy range of 200~keV to 1400~keV~\cite{Nucl.datasheet.152}. The time resolution of the strongest cascade in \isotope[152]{Gd} with a 779~keV ($3^-_1 \rightarrow 2^+_1$) populating and a 344~keV ($2^+_1 \rightarrow 0^+_1$) depopulating transition and a short lifetime of 46.3(39)~ps~\cite{Nucl.datasheet.152} is used as a benchmark during the optimization process.\par
\begin{figure}[t]
\centering
    \includegraphics[width={230.00bp},height={240.00bp}]{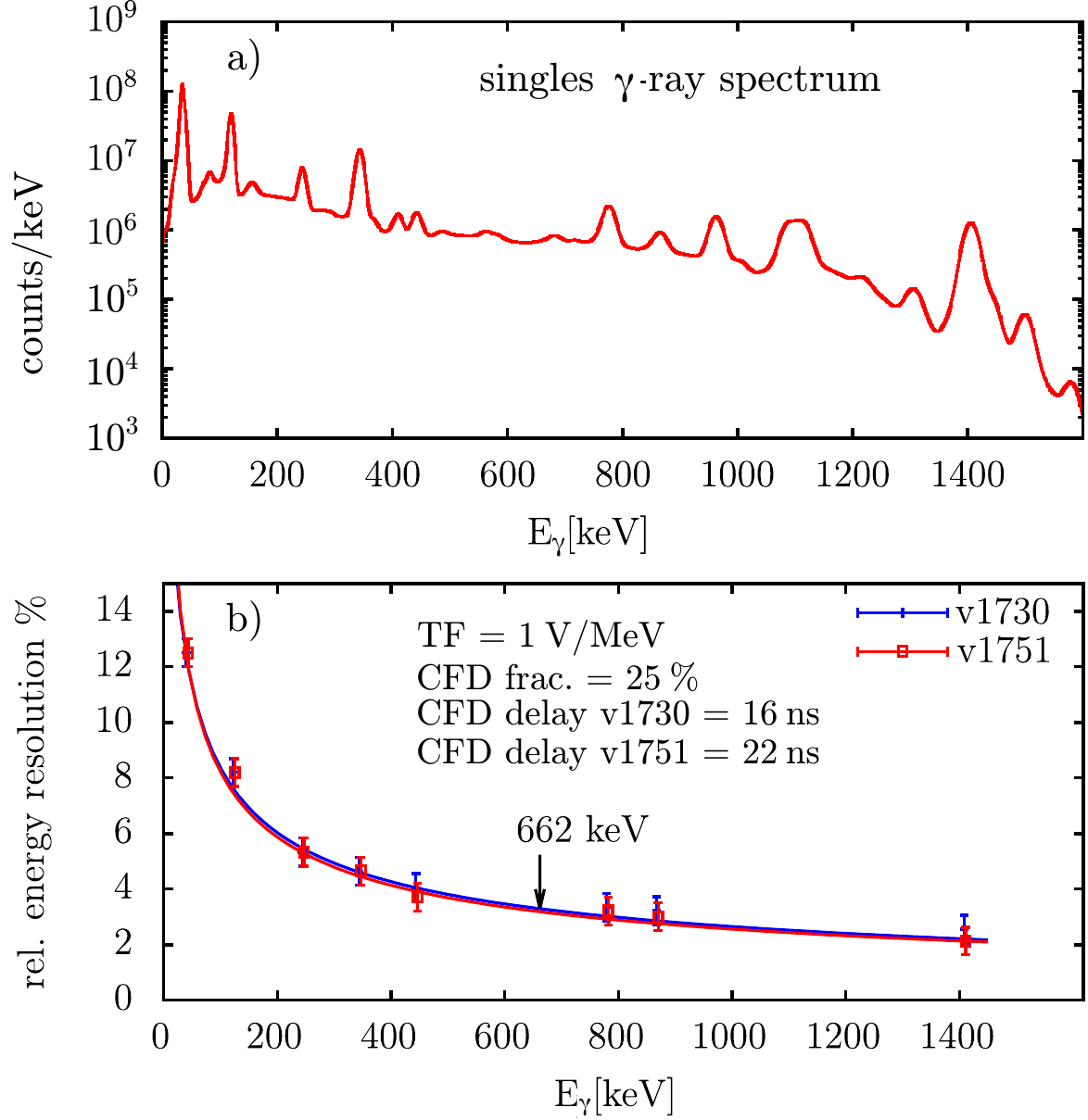}
    \caption{(a) A singles $\gamma$-ray spectrum of the \isotope[152]{Eu} $\gamma$-ray source for the V1730 digitizer. (b) Relative energy resolutions of both digitizers shown over an energy range between approximately 100 and 1400~keV, demonstrating a decreasing behavior with respect to $1/\sqrt{E_{\gamma}}$.}
    \label{fig:spectrumReso}
\end{figure}

In Fig.~\ref{fig:spectrumReso}a, an exemplary singles $\gamma$-ray spectrum for the V1730 digitizer is given and Fig.~\ref{fig:spectrumReso}b shows the relative energy resolution $\Delta$E/E [\%] for both digitizers in an energy range between 40 and 1400~keV. The relative energy resolution is optimized through gain matching of the electron multiplication of the PMTs~\cite{LESKOVAR.PMTgain}. It amounts to  3.20(27)~\% and 3.39(27)~\% for the V1751 and the V1730 at a $\gamma$-ray energy of 662~keV, which is commonly used as a benchmark for energy resolution, respectively.

\subsection{Description of dedicated measurement approaches}
\label{ssec:approaches}
The crucial properties of implemented interpolating CFD algorithms in context of fast-timing techniques are the time resolution and time walk. Both properties are depending on the shaping parameters "CFD delay", "CFD fraction" and the input signal amplitude. To classify the time resolution and time walk characteristics of the digitizers, several measurements of time periods between 10 and 24 hours with different parameter settings have been conducted. Different parameter settings were used, with two of the three parameters kept fixed, while the third was varied.\\
\indent 
The voltage applied to the PMTs was adjusted to align the input signal amplitudes of the different channels to the amplitude of the 1408~keV $\gamma$-ray emitted by the \isotope[152]{Eu} source. The transfer function (TF) describes the relationship between the output voltage of the PMT and energy of the incident $\gamma$-ray. It is expressed in V/MeV and serves as a measure for the input amplitude. The TFs in the present measurements are nearly linear, as the PMTs are operated in a low voltage range of 750 to 1500~V.\\
\indent In order to evaluate the intrinsic time jitter of the digital CFDs, an additional measurement was conducted using a split signal configuration with a single LaBr detector, as described in previous studies~\cite{LIPSCHUTZ.splitsignal,warburton.timingAlg}. The anode signal from the LaBr detector was split using a T-adapter and the two resulting signals were fed into separate signal inputs of the digitizers, as illustrated in Fig.~\ref{fig:setup}b. Two digitally interpolated timestamps for the same detector pulse are acquired. Time difference distributions derived from these timestamps are free of energy-dependent time walk, lifetime effects and nearly free of other systematic effects. This approach allows for a determination of the intrinsic time jitter of the digital CFDs as a function of the input signal amplitude only effected by the SNR (see Sec.~\ref{sec:cfd}).\par

\subsection{Data Analysis of the measurements}
\label{ssec:analysis}
The data sets obtained from the fast-timing setup with four LaBr scintillators were analyzed using the ftSOCO code~\cite{ftsoco}, which was designed for the analysis of digital fast-timing data.
Fast-timing events were generated requiring exactly 2 LaBr hits within a coincidence window of 20~ns, which is sufficient for the purpose of this work (see Fig.~\ref{fig:zoom}).
If this multiplicity condition is not used in multi-detector arrays, many time-uncorrelated background events (especially inter-detector Compton scattering events) are incremented. As addressed in Sec.~\ref{sec:code}, small runtime differences in the arrival times of the detector signals are corrected by applying a constant timestamp offset for every detector-detector combination.\par
For all measurements, the centroid shift method, as described in Sec.~\ref{ssec:two-diff-appr}, was used to determine the TW characteristic~\cite{REGIS.redyyTW.multiplexed,REGIS.symmetricCube}.
Each analysis for different CFD parameters or transfer functions was performed using identical energy gates on the transitions of interest to allow for comparable results.\par
\begin{figure}[t]
\centering
    \includegraphics[width={250.00bp},height={128.00bp}]{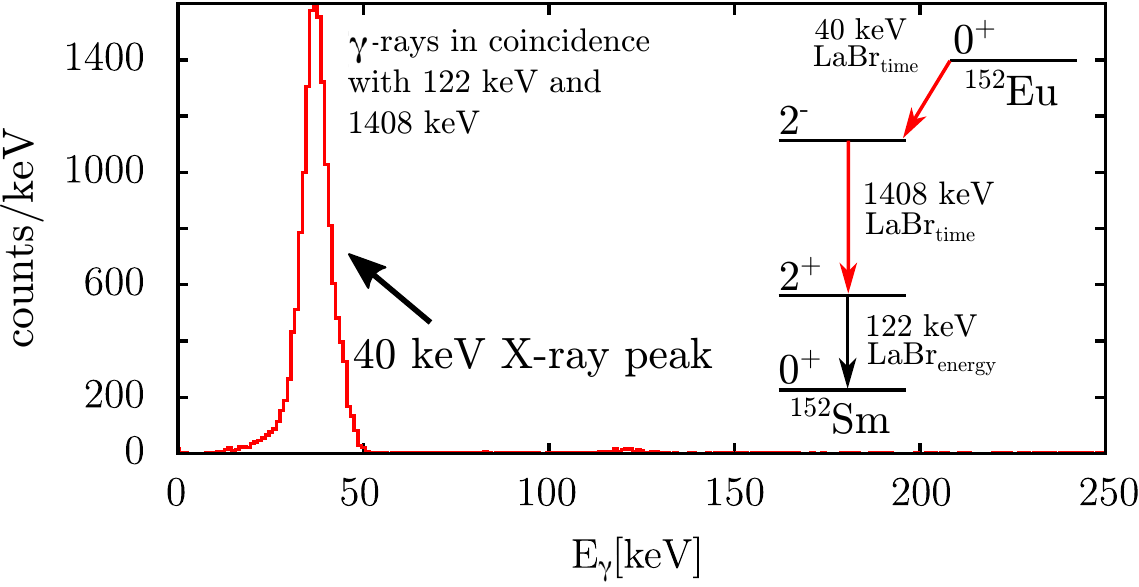}
    \caption{Coincidence spectrum of the threefold coincidences with gates on the 122~keV and 1408~keV $\gamma$-rays from the \isotope[152]{Eu} source. It is evident that there is relatively little background under the 40~keV X-ray peak. On the right, the decay scheme from \isotope[152]{Eu} to \isotope[152]{Sm}, with the transitions used for timing (40~keV and 1408~keV) marked in red, is depicted.}
    \label{fig:threefold}
\end{figure}

To investigate the low energy time walk of the digitizers, a data point below 100~keV is needed. This is achieved by utilizing threefold coincidences of the cascade 40~keV - 1408~keV - 122~keV (X-ray~$\rightarrow~2^-~\rightarrow~2^+~\rightarrow~0^+$) in \isotope[152]{Sm}. The 40-keV K-X-ray is emitted after the electron-capture decay of \isotope[152]{Eu} to \isotope[152]{Sm} within femtoseconds, which is significantly shorter than the lifetimes of the subsequent excited states in \isotope[152]{Sm}. The detection of the 40-keV K-X-ray indicates the moment of population and acts as a nearly prompt feeding transition for the $2^-_1$ state in \isotope[152]{Sm}~\cite{REGIS.redyyTW.multiplexed}. Usually, a high-resolution high-purity Germanium detector is used to apply an energy gate on the 122~keV transition. For the setup used in this work, the pairs of face-to-face detectors (same color in Fig.~\ref{fig:setup}) were used for timing coincidences, while the neighbouring detectors were used for the additional energy gate on the 122~keV transition. A coincidence spectrum and decay scheme for the threefold LaBr coincidences with energy gates on the 122~keV and 1408~keV $\gamma$-rays is shown in Fig.~\ref{fig:threefold}. Note, that not all measurements had sufficient statistics in the X-ray region to allow for the use of this approach.\par
The data obtained from the split signal setup (Fig.~\ref{fig:setup}b) was analyzed using a coincidence window of 2~ns. A double gating method was employed, where the same energy gate was applied to both channels, since the same detector pulse was fed into both channels. The resulting time difference distributions provided insight into the precision of time difference determination between signals of equal energy with a consistent time difference of zero. A range of amplitude between 100 and 1000~mV was scanned using this method.\par
In some time difference distributions, periodic binning artefacts with low picosecond periods have been observed. These artefacts are possibly a result of the fine timestamp T$_{\textrm{fine}}$ being reported as a 10-bit integer by the FPGA, resulting in a precision of T$_{\textrm{sample}}$/1024. The resulting precision of T$_{\textrm{fine}}$ is 1.953125 $\approx$ 2~ps in the exemplary case of the V1730~\cite{CAEN.compass}. Most likely, this approximation causes an effect on the time differences, which shows up as a periodic oscillations and additional beat frequencies in the time difference distributions. The binning artefacts have no influence on the centroid shift analysis and were addressed by randomly blurring a calculated time difference over a two-picosecond window.\par

\section{Results of the systematic investigation of timing characteristics}
\label{sec:results}
Modern fast-timing system are able to achieve time resolutions between 300-400~ps for the benchmark cascade 779-344~keV~\cite{REGIS.yy-fasttiming}. For the time walk characteristic a mostly linear and flat progression is desired with variations in a range of 100~ps or less~\cite{REGIS.redyyTW.multiplexed}. 
To sufficiently cover the energy range that is significant for lifetime measurements in standard nuclear structure investigations, the amplitude input range of a timing system should be maximised, but it should at least extend up to $\gamma$-ray energies of 1.5~MeV~\cite{REGIS.yy-fasttiming, Esmayl.fasttiming2, Harter.fasttiming, karayon.fasttiming, knafla.fasttiming, AEsmayl.fasttiming, vkarayon.fasttiming}.
In the following sections, the results of our measurements obtained using different parameter combinations are presented and discussed. Specifically, the time resolution and time walk of the digital CFDs in comparison to the established analog setup concerning the above expectations are evaluated.
\par

\subsection{Time jitter considerations}
\begin{figure}[t]
\centering
    \includegraphics[width={230.00bp},height={250.00bp}]{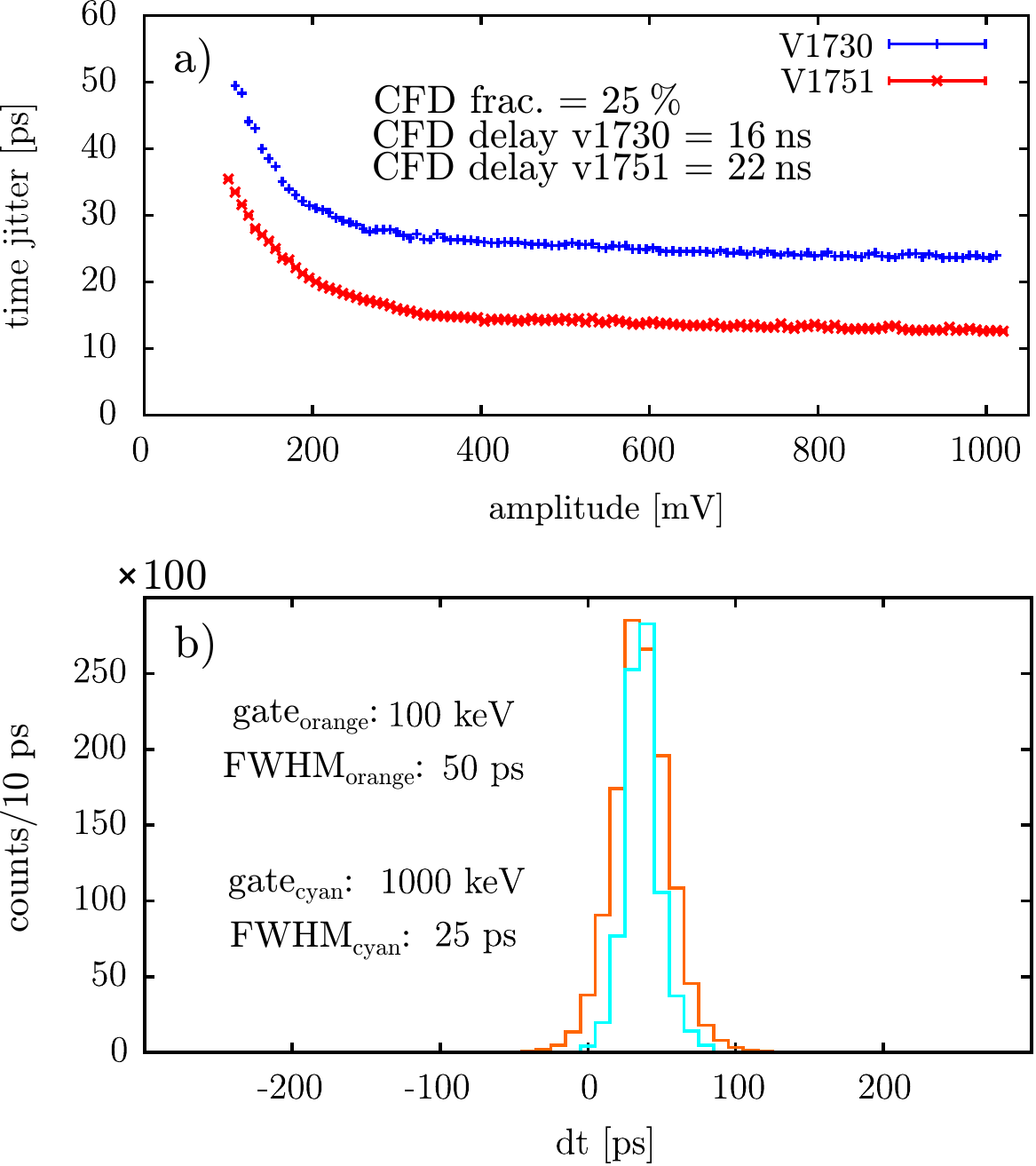}
    \caption{(a) The CFD time jitter as extracted from the split signal measurement is plotted against the signal amplitude in a range from approximately 100~mV to 1000~mV. (b) Two exemplary time difference distributions of the split signal setup with a V1730 digitizer. The double gates are set on 100~keV with a FWHM of 50~ps and on 1000~keV with a FWHM of 25~ps. For details see text.}
    \label{fig:jitter}
\end{figure}

 In Figure~\ref{fig:jitter}a, the progression of the time jitter as obtained from the split signal setup is illustrated. The influence of the SNR on the time jitter is observed in the low amplitude region below 400~mV where a decrease in SNR results in a corresponding increase in time jitter (see Sec.~\ref{sec:cfd}). The time jitter saturates around 25~ps and 13~ps for the selected channels of the V1730 and V1751, respectively, with a difference of 12~ps. Figure~\ref{fig:jitter}b depicts time difference distributions for two different double-gates (described in Sec.~\ref{ssec:analysis}) at 100 and 1000~mV for the V1730. The FWHMs obtained were 50 and 24~ps, respectively, and the uncertainties can be considered negligible.\par
\begin{figure}[t]
\centering
    \includegraphics[width={240.00bp},height={130.00bp}]{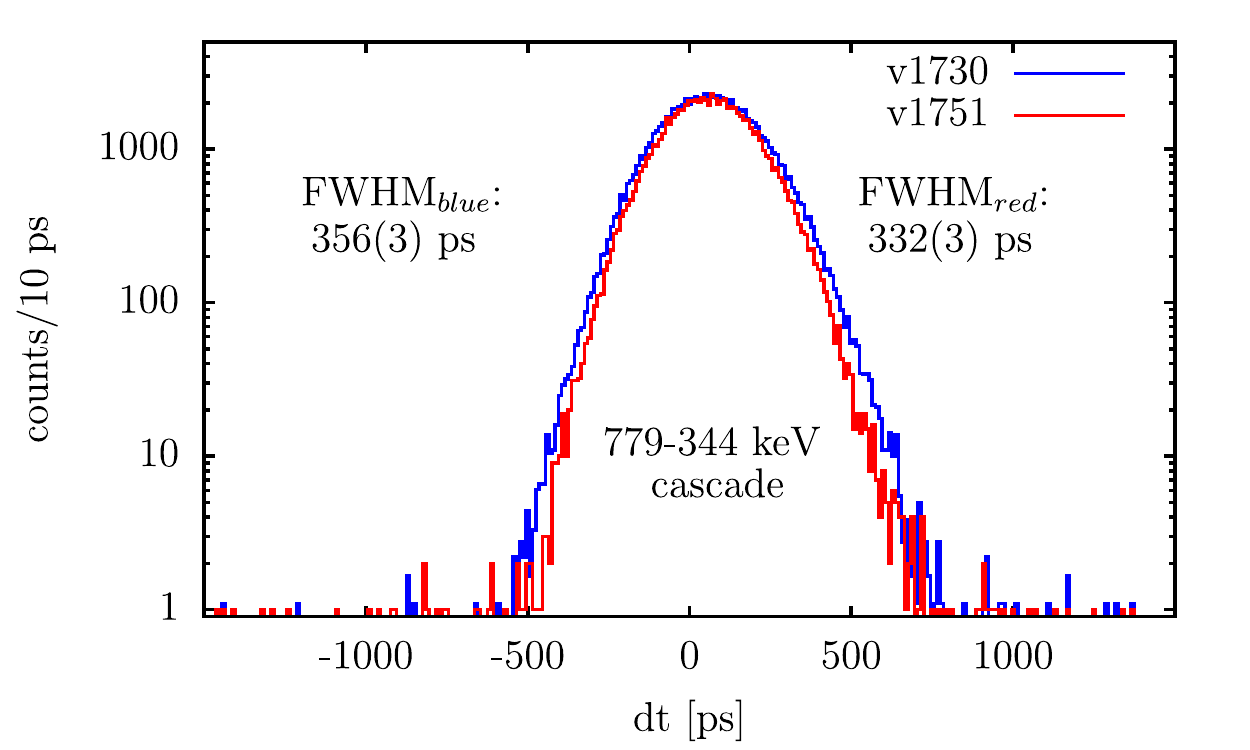}
    \caption{Exemplary time difference distributions of the 779-344~keV cascade for both digitizers. The blue distribution is obtained from the V1730 and has a FWHM of 356(3)~ps. The red distribution is obtained from the V1751 and has a FWHM of 332(3)~ps. The parameters as displayed in Tab.~\ref{tab:params} where used to generate these distributions.}
    \label{fig:centroid}
\end{figure}
The time difference distributions in Fig.~\ref{fig:centroid}, obtained from the investigated digitizers, reveal that the time resolutions between the two digitizers differ by about 25~ps for the considered channel combinations. This discrepancy in time resolution can be attributed to variations in time jitter between the digitizers. The 24~ps difference in time resolution for the investigated channel combinations of the digitizers corresponds to twice the difference in time jitter (12~ps) between the considered channel combinations of the boards. This approximation holds if all channels of one board have comparable time jitters and if the CFD contribution to the time resolution corresponds to the summed time jitters of the used channels.\par

\subsection{Influence of the CFD algorithm on the time resolution}
\begin{figure}[t]
\centering
    \includegraphics[width={230.00bp},height={200.00bp}]{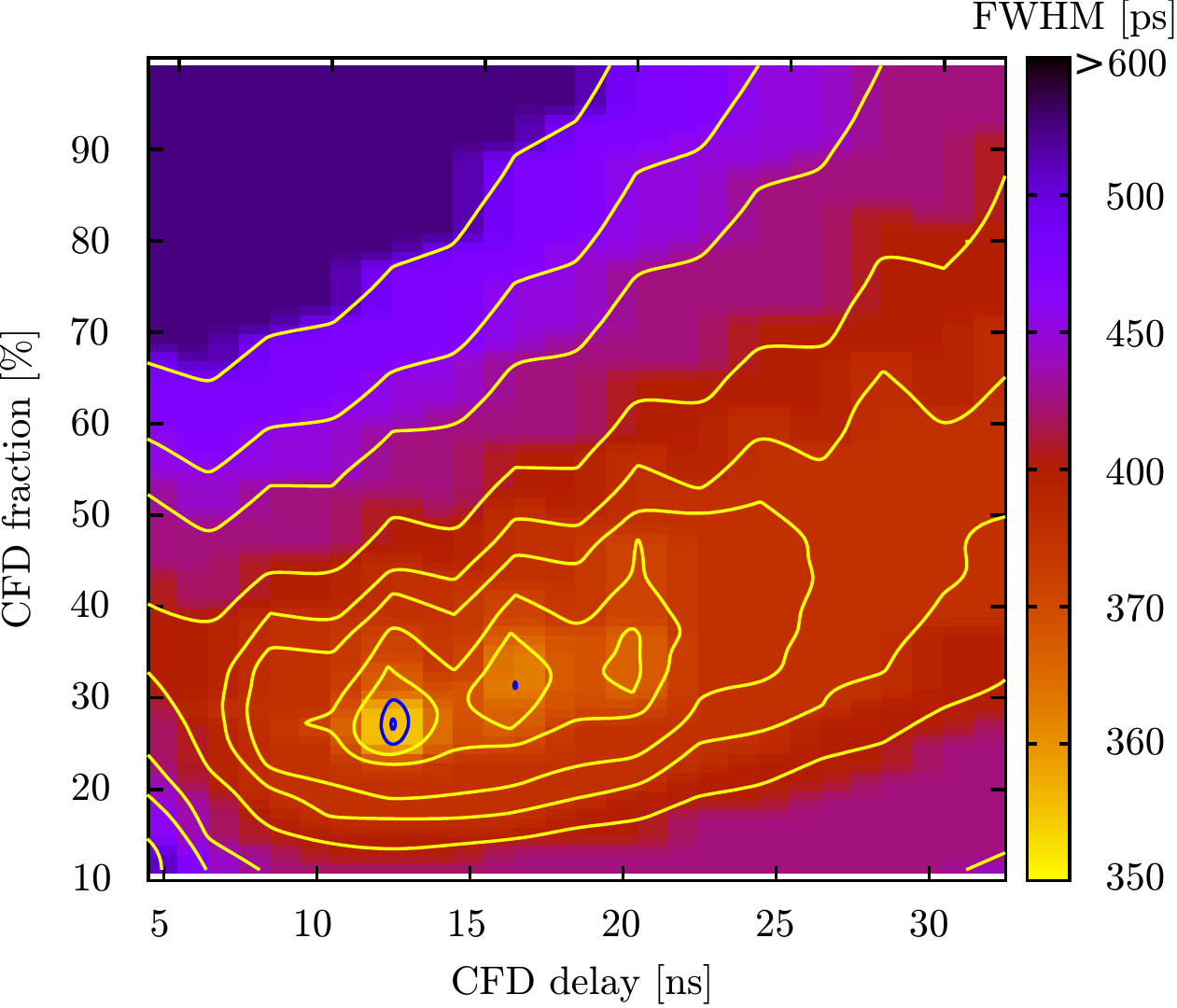}
    \caption{CFD parameter scan in terms of time resolution (FWHM) of the benchmark cascade in the CFD delay - CFD fraction parameter space. Contours are given for selected heights. The best time resolutions are delivered by parameters around CFD delay = 12~ns and CFD fraction = 27~\%, indicated by the blue contours. The transfer function used for the measurement of this dataset was 1~V/MeV.}
    \label{fig:param-heatmap}
\end{figure}
During this study, the CFD algorithm from the digitizers in use was re-implemented in an offline analysis and applied to a dataset of wave traces recorded with a V1730 to get a comprehensive overview of the CFD parameter space. The offline implementation is more flexible than the online implementation since the CFD fraction is not limited to four values. A section of the CFD parameter space was scanned using the offline CFD implementation where the CFD delay was selected between 4 and 32~ns in 2~ns steps and the CFD fraction was selected between 10\% and 100\% in 2~\% steps. The heatmap shown in Fig.~\ref{fig:param-heatmap} represents the surface of the time resolution of the benchmark cascade. The yellow area around CFD delay~=~14~ns and CFD fraction~=~27~\%, marked by the blue contours, yields the best time resolution values, which is largely confirmed by the online measurements, which is discussed in the next section. 
With the offline parameter scan, a preliminary visualization of the parameter space was generated to reduce the amount of required experimental measurements to about 10~\%, and the parameters for the online measurements could be selected accordingly.\par
\begin{figure}[t]
\centering
    \includegraphics[width={230.00bp},height={371.00bp}]{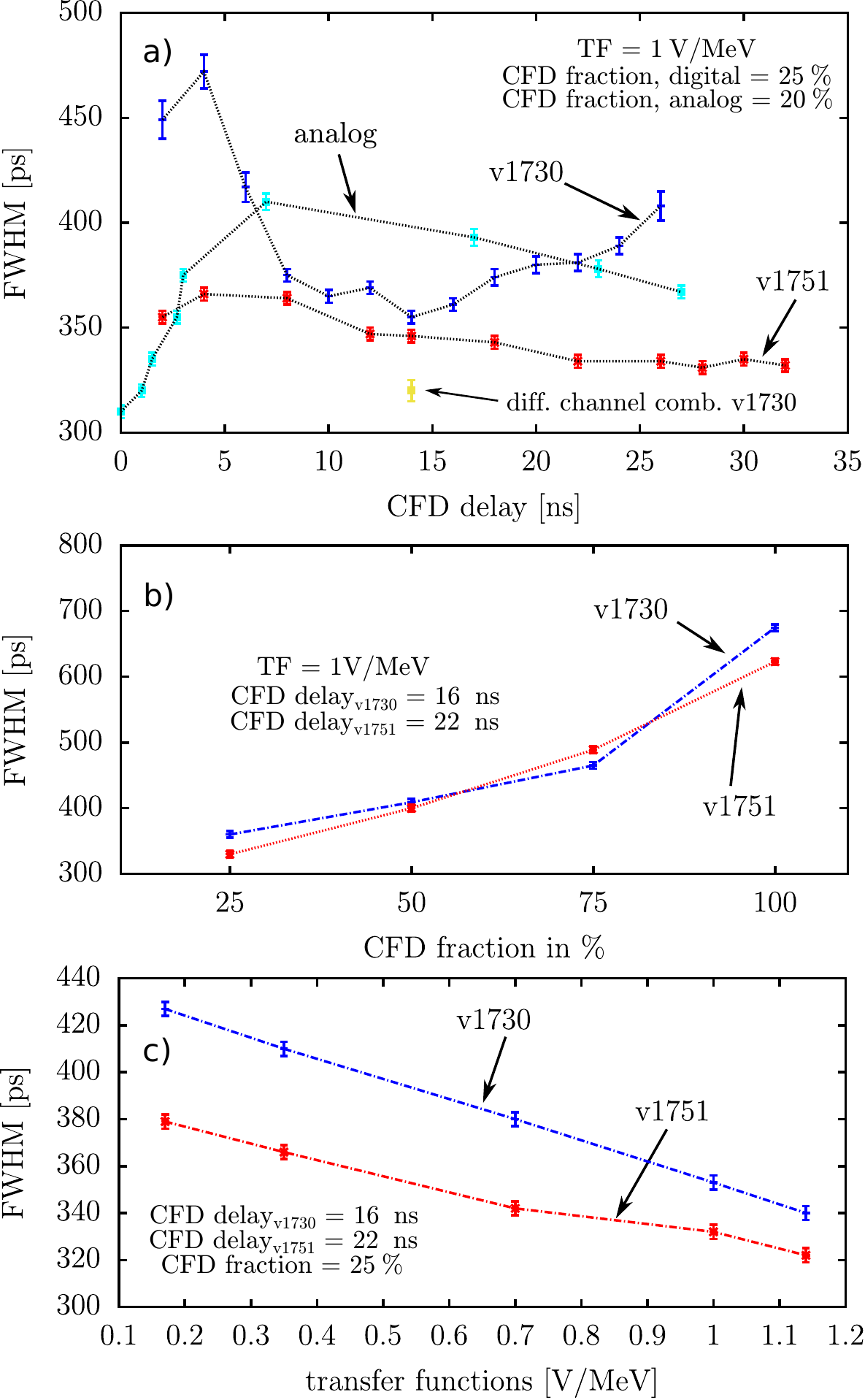}
    \caption{Time resolutions (FWHM) of both digitizers versus (a) CFD fraction, (b) transfer functions and (c) CFD delay in comparison with the analog measured time resolution using the CFD model ORTEC 935. For all plots, the time resolution corresponds to the FWHM of the benchmark cascade 779 - 344~keV obtained from a \isotope[152]{Eu} source. The parameters of the CFD and the TF are given in each plot. The analog data points are taken from Ref.~\cite{REGIS.redyyTW.multiplexed}. The dashed lines serve to guide the eye.}
    \label{fig:timeVSall}
\end{figure}

In the following, the results of the optimization process aimed at identifying the optimal combination of CFD delay, CFD fraction, and detector signal amplitude are presented and discussed in terms of their impact on the time resolution of the digital CFDs. In the plots presented in Fig.~\ref{fig:timeVSall}, the constant parameter sets utilized during the single measuring series are given. In Fig.~\ref{fig:timeVSall}a, the time resolution as a function of the CFD delay parameter is presented for both digitizers. The V1730 exhibits a minimum in time resolution at a CFD delay value of 14~ns, which fits to the prediction of the offline parameter scan, see Fig.~\ref{fig:param-heatmap}. In contrast, the V1751 does not display a well-defined minimum over the range of tested CFD delay values. However, it generally exhibits a decreasing trend in time resolution when increasing the CFD delay across the majority of the tested range, remaining relatively constant above 22~ns. For comparison, in the same plot the dependence of the time resolution of the delay parameter of an analog setup using a CFD model ORTEC 935 is depicted and it moves in a comparable range~\cite{REGIS.redyyTW.multiplexed}. The CFD delay parameters used for the further optimization process were 16~ns and 22~ns for the V1730 and the V1751, respectively.\par

In Fig.~\ref{fig:timeVSall}b, the time resolution as a function of the CFD fraction parameter is presented for both digitizers. The influence of the CFD fraction on the time resolution is significant. As the CFD fraction increases from 25~\% to 100~\%, the time resolution degrades dramatically. For both modules the smallest available CFD fraction (25~\%) provided by the firmware~\cite{CAEN.dpppsd.2022} yields the best time resolution. Other studies suggest, that CFD algorithms and analog CFDs with CFD fractions around 10~\% provide better time resolutions than a fraction of 25~\%~\cite{Nakhostin2014,MCDONALD.cfdfractions,Hyman.cfdfractions}. Based on the offline parameter scan in Fig.~\ref{fig:param-heatmap}, this cannot be confirmed by this study.\par
In Fig.~\ref{fig:timeVSall}c, the dependency of the time resolution on the TF is shown. As expected from another study~\cite{BHATTACHARYA.PMT-voltages}, a strong nearly linear dependence of time resolution on the TF is observed for both digitizer modules. The time resolution improves with increasing TF. The underlying reasons for this behavior are discussed in Section~\ref{sec:cfd}. It should be noted that the impact of the PMT voltage and signal amplitude on the time resolution is non-negligible. Factors such as the number of photoelectrons collected~\cite{FALLULABRUYERE.CFDformel,Aykac-PMT} and transit time spread~\cite{DELABARRE.tts}, play a significant role in the improvement of the time resolution of PMT systems. Therefore, it is likely that the PMT effects are dominating the influence of the digital CFDs here.\par
\begin{figure}[t]
\centering
    \includegraphics[width={250.00bp},height={130.00bp}]{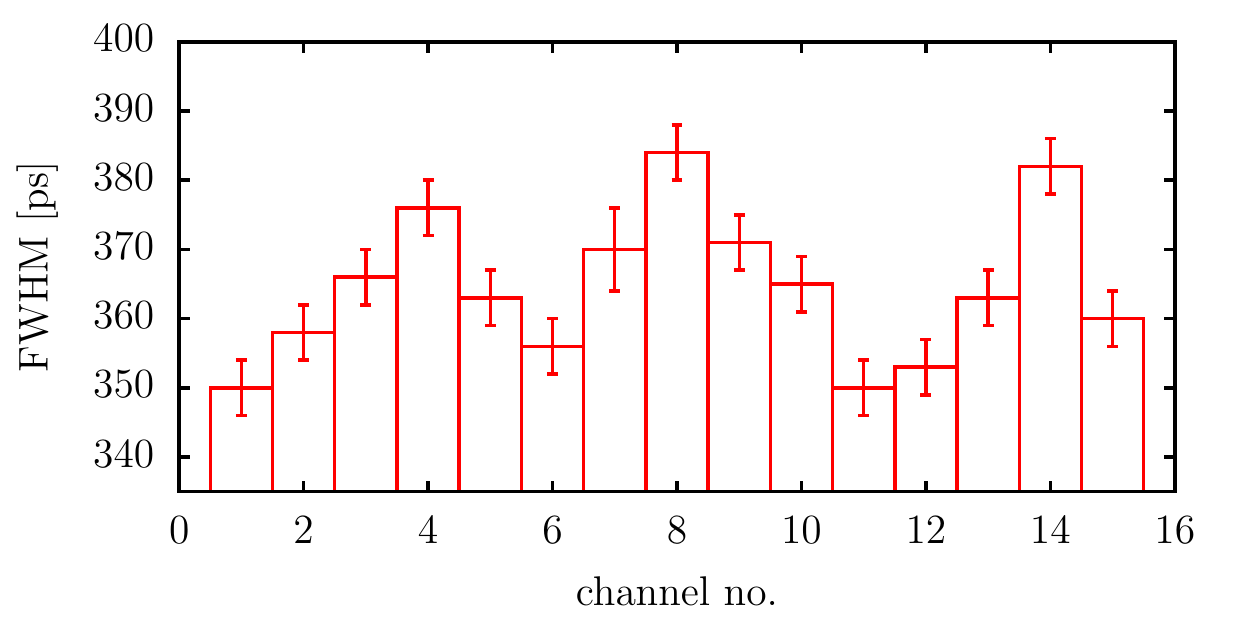}
    \caption{Time resolutions of the different channels of a specific V1730 digitizer module in relation to channel 0. These differences are board specific and have to be quantified separately for every board. For details see text.}
    \label{fig:resvsChannel}
\end{figure}

In conclusion, the optimal parameters for the CFD settings and TF with respect to the time resolution as determined in this study are provided in Tab.~\ref{tab:params}, along with the corresponding time resolution values for both digitizers, which are considered as the optimal values obtained in this study.
It is noteworthy, that considerably different time resolutions were observed while using a different channel combination than the one studied in this research, which remained unchanged during the measurements. Time resolution values of about 320~ps were obtained with the same TF, detectors and CFD settings for the V1730 but utilizing a different channel combination. Apparently, the time resolution shows significant differences between the different input channels of the digitizers. In Fig.~\ref{fig:resvsChannel}, a comparison of the time resolutions for a single V1730 digitizer across different channel combinations in relation to channel 0 is presented. One detector was consistently connected to channel 0, while the other one was cycled through all other channels. These differences are board and channel specific and have to be quantified separately for every board. Currently, there is no definitive explanation for this behavior. However, it is plausible that it is related to variations in internal reference clock runtime within the digitizer and differences in the resolution of a single sample among the different channels.\par

\begin{table}[t]
    \caption{The best parameter sets for the digitizer modules under investigation applicable for fast-timing in an energy range from around 100 to 1400~keV. The time resolutions are given for the PMTs and digitizer channel combinations used for the investigations and can deviate from the time resolutions of other channel combinations and PMTs. For details see text.}
    \centering
    \begin{tabular}{cccc}
     parameter & V1730 & \multicolumn{2}{c}{V1751}\\
     \hline \hline
       & & opt. time resol. & opt. TW \\
     \hline 
     CFD delay & 14~ns & 22~ns & 22~ns \\
     CFD fraction & 25~\% & 25~\% & 25~\%  \\
     TF & 1~V/MeV & 1~V/MeV & 0.7~V/MeV \\
     \hline
     time resolution & 356(3)~ps & 332(3)~ps & 342(3)~ps
    \end{tabular}
    \label{tab:params}
\end{table}

\subsection{Results of the time walk investigations}
\begin{figure}[t]
  \centering
      \includegraphics[width={224.00bp},height={419.00bp}]{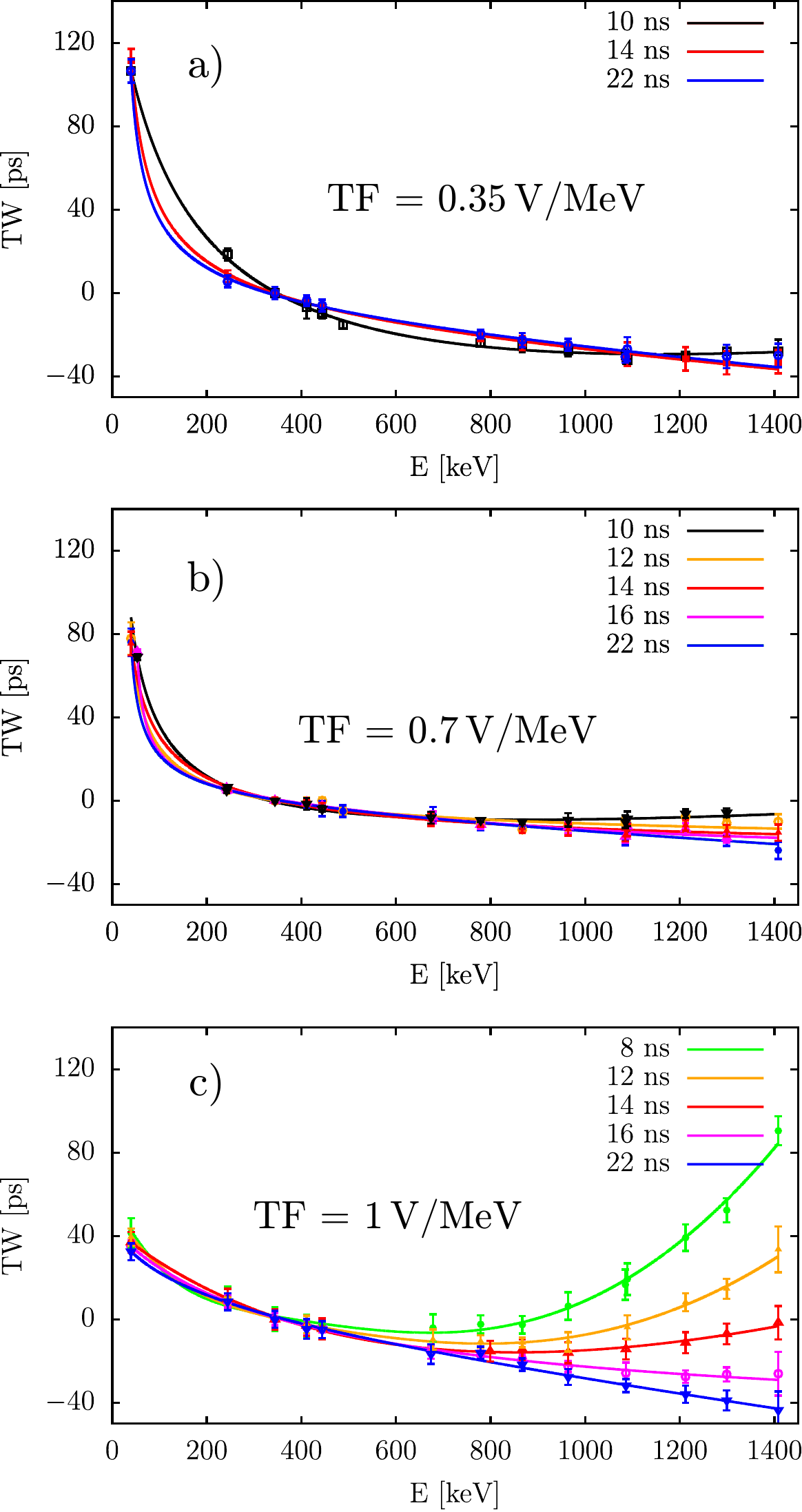}
    \caption{Time walk characteristics of the V1730 digitizer for different TFs in dependence of the CFD delay.}
    \label{fig:tw.V1730}
\end{figure}%
\begin{figure}[t]
  \centering
   \includegraphics[width={224.00bp},height={419.00bp}]{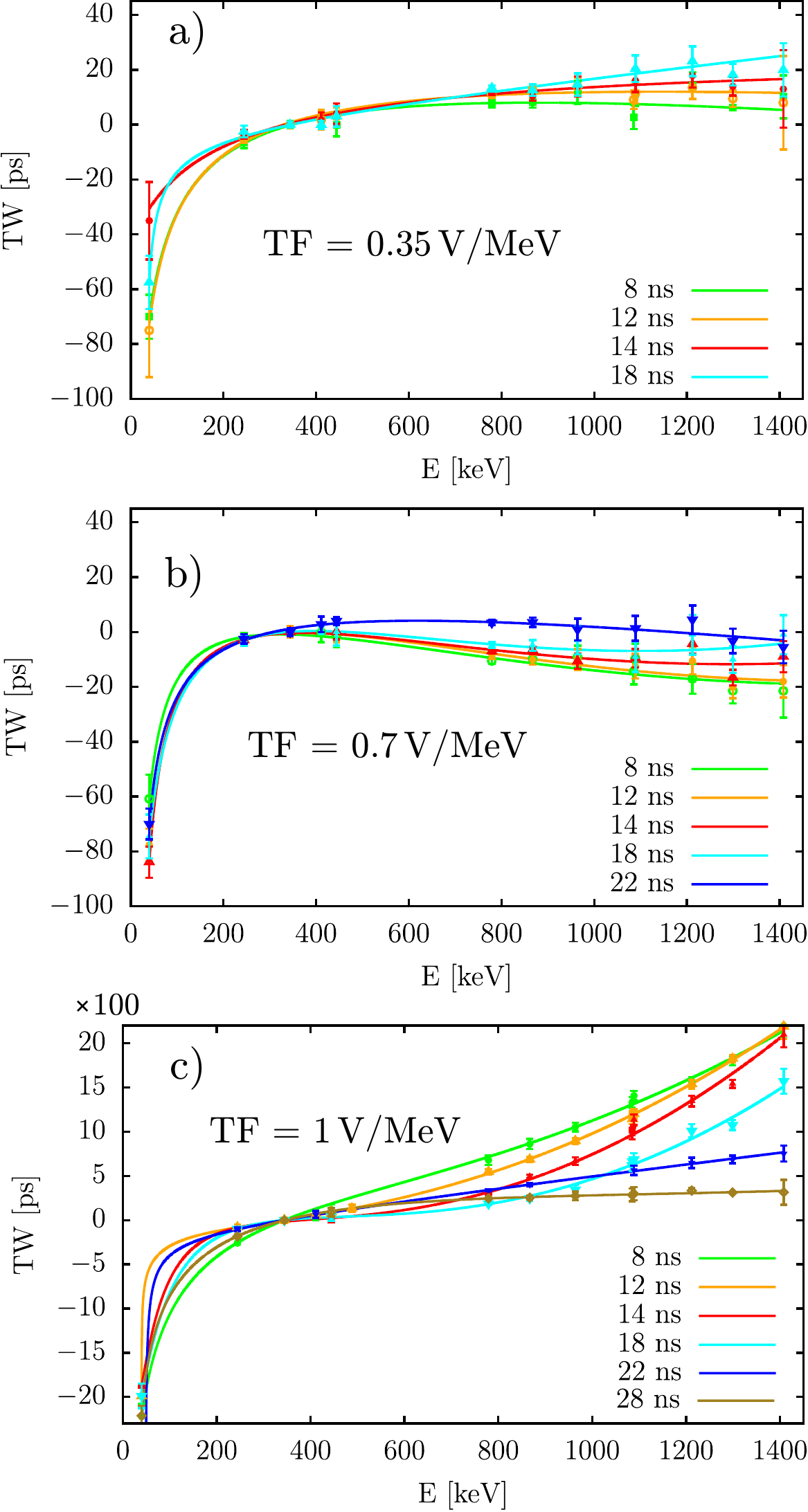}
    \caption{Time walk characteristics of the V1751 digitizer for different TFs in dependence of the CFD delay. Note the different y-scale in c).}
    \label{fig:tw.V1751}
\end{figure}
Next, we will focus on the investigations of the time walk characteristics of the implemented interpolating CFD algorithms employed in the digitizer models under consideration. The time walk calibration is done according to Eq.~\ref{eq:tau} using:
$$
TW(\textrm{E}_1,\textrm{E}_2) = C_D(\textrm{E}_1,\textrm{E}_2) - T_0 - \tau 
$$
The time walk can be calibrated across a 40-1408~keV $\gamma$-ray energy range using a \isotope[152]{Eu} $\gamma$-ray source decaying through multiple feeder-decay cascades connecting different intermediate states with well known lifetimes. The TW function is generally defined by 
\begin{equation}
TW(E) = \frac{a}{\sqrt{E-b}}+c+dE+eE^2.
\end{equation}
The TW calibration procedure is similar to the prompt response difference (PRD) calibration of an analog setup~\cite{REGIS.redyyTW.multiplexed,REGIS.CFD}, which is detailed in Refs.~\cite{REGIS.redyyTW.multiplexed, REGIS.symmetricCube}. TW and PRD have the same properties and are connected by~\cite{REGIS.symmetricCube}
\begin{equation}
    TW(\textrm{E}_1,\textrm{E}_2) = TW(\textrm{E}_2) - TW(\textrm{E}_1) = \frac{PRD(\textrm{E}_1,\textrm{E}_2)}{2}.
\end{equation}

For the investigation of the TW characteristics, the CFD fraction was kept constant at 25~\% and the TW was investigated with respect to different CFD delays and TFs. Figures~\ref{fig:tw.V1730} and \ref{fig:tw.V1751} show the time walk characteristics of both digitizers as a function of the $\gamma$-ray energy.
To be able to describe the strong leading edge component~\cite{REGIS.redyyTW.multiplexed,Paulus.CFD} in the low energy region in case of the V1751 with TF = 1~V/MeV (Fig~\ref{fig:tw.V1751}c), the TW function was generalized to
\begin{gather*}
    TW(E) = a(E^m-b)^{-(1/q)}+c + dE + eE^2, \\  m \in \{1,2\}, \; q \in \mathbb{Q}^+.
\end{gather*}
\par
Figures~\ref{fig:tw.V1730} a) to c) show the evolution of the time walk characteristics of the V1730 digitizer in dependency of the CFD delay and the TF. The behavior of the TWs of the V1730 in the low amplitude range is characterized by a downward trend and in the high amplitude range, from 500~keV upward, it shows a drop, ascent, or level behavior in dependence of the CFD delay. The investigation showed, that the CFD delay has nearly negligible influence in the low energy region. However, the progression of the high-energy time walk is dominated by the influence of the CFD delay parameter. The maximum TW difference in the energy range between 240~keV and 1300~keV lies below 100~ps for all TFs and specially below 25~ps for a TF = 0.7~V/MeV depending on the CFD delay. Moreover, Fig.~\ref{fig:tw.V1730}c shows nearly the same shape and magnitude of time walk in dependence of the CFD delay time observed using ORTEC 935 with same TF = 1~V/MeV (Fig.~8 in Ref.~\cite{REGIS.redyyTW.multiplexed}). This systematic is identical to the behavior of the analog ORTEC 935 with comparable time resolution and TW range.\par
Analogous, the TW behavior of the V1751 is shown in Fig.~\ref{fig:tw.V1751} a) to c). A mirrored behavior is observed compared to the TW characteristics of the V1730. The TW curves all show increasing behavior in the low energy range independent from the CFD delay. The contrasting time walk behavior of the V1730 and the V1751 suggests that the CFD algorithm in the V1751 is internally inverted compared to the one in the V1730. The TW curves presented in Ref.~\cite{DasPMT.CFD}, where a CFD model ORTEC 584 was used, show a similar behavior. The ORTEC 584 performs the shaping procedure in a sequence like described in Eq.~\ref{eq:cfd2}~\cite{ORTEC.584}. For a negative input signal, this leads to a negative slope at the zero crossover and thus an inverted time walk characteristic. A closer look at the manual of the firmware of the V1751 shows, that the DPP-PSD firmware of the V1751~\cite{CAEN.dpppsd.2022} indeed uses a similar inverted CFD algorithm as the ORTEC 584.\par
As visible in Fig.~\ref{fig:tw.V1751}a, the CFD delay shows its strongest influence in the high energy range from 500~keV upward, like in the V1730. In the low energy range, the curves show a similar behavior nearly independent from the CFD delay. In Fig.~\ref{fig:tw.V1751}b, the TW curves are plotted for delays starting from 8~ns to 22~ns with a TF of 0.7~V/MeV. The curves with CFD delays smaller than 22~ns, have two inflection points, different to all other curves before. For CFD delay = 22~ns, there is a transition in the TW characteristic from a convex to a concave shape especially for energies higher than 300~keV. The progression of the TW curve reverts to the typical characteristics, showing only a single inflection point. This behavior is even more clearly observable in the progression of the TW curves for a TF = 1~V/MeV in Fig.~\ref{fig:tw.V1751}c. An explanation for this transition behavior needs further investigation of the CFD algorithms and the digitizer modules and can not be given at the state of this work.\par
The time walk of the CFD algorithm of the V1751 in the case of the highest TF = 1~V/MeV increases dramatically up to above 1.5~ns. The maximum time walk between 200~keV and 1300~keV in the case of TF = 0.35~V/MeV and TF = 0.7~V/MeV lies below 25~ps depending on the CFD delay. The smallest maximum range of the TW throughout the considered energy range between about 240 and 1300~keV is found for a CFD delays between 14 and 22~ns for both modules.\par

\section{Summary and conclusion}
\label{sec:summary}
In this work, the digitizer modules V1730 and V1751 by CAEN S.p.A. with sample rates of 500~MS/s and 1~GS/s, respectively, were investigated with regard to the fast-timing properties of the implementations of the digital real-time interpolating CFD algorithms. Both modules provide linear interpolation algorithms to determine the timestamp of an incoming energy pulse from a connected fast-timing scintillator within sub-sample-period picosecond precision. To evaluate the fast-timing characteristics of the digital CFDs, a fast-timing setup, consisting of four LaBr detectors was constructed to test both digitizer modules. A standard \isotope[152]{Eu} time walk calibration source was used to investigate the time resolution and time walk over a $\gamma$-ray energy range of 40 - 1408~keV.\par
Our study showed, that the timing properties of the digital CFDs of these digitizers are comparable to the characteristics of the analog model ORTEC 935. The determined time resolutions of about 350~ps or lower and time walk characteristics demonstrate that both digitizers are suitable for fast-timing with modern standards.
The time walk characteristics of the V1730 has very small maximum ranges of around 25~ps in the considered energy range throughout all measured TFs and is comparable to the characteristics of analog modules like ORTEC 935. The time walk characteristics of the V1751 is comparable to the analog module ORTEC 584 for low TFs up to 0.7~V/MeV but shows limitations in usability for higher TFs.\par
Using a V1730 with a input dynamic range of $\pm$ 2~Vpp and a TF of 1~V/MeV, it is possible to detect $\gamma$-rays up to about 4~MeV. In case of a V1751 with an input dynamic range of $\pm$ 1~Vpp and a TF of about 0.7~V/MeV a comparable $\gamma$-ray energy range is realized.\par
In conclusion, the V1730 was found to be a highly effective and user-friendly instrument for achieving modern fast-timing standards. Although the V1751 possesses a smaller input dynamic range and a larger maximum range of time walk for high TFs, it remains a valuable tool for fast-timing for lower $\gamma$ energy ranges.
Both the investigated digitizers and similar modules offer high-quality performance in fast-timing measurements, while remaining easy to operate even for large fast-timing arrays.\par
Both modules have potential for further improvement, e.g. expanding the linear interpolation algorithm to a cubic one or providing more selectable CFD fraction values. However, other digitizers feature larger FPGAs that accommodate the proposed improvements, but further studies are needed to fully explore the potential benefits. Still, using digitally implemented linear interpolating CFD algorithms, fast-timing experiment without significant compromises with regard to time resolution and time walk behavior are possible.

\section{Declaration of competing interests}
The authors declare that they have no known competing financial interests or personal relationships that could have appeared to influence the work reported in this paper.

\section{Acknowledgement}
J.-M. R. and M. L. acknowledge the Deutsche Forschungsgemeinschaft for support under grant No. JO 391/18-1. A. E. wants to acknowledge the support of BMBF Verbundprojekt 05P2021 (ErUM-FSP T07) under grant No. 05P21PKFN1. All authors want to acknowledge and especially thank CAEN S.p.A. for the free loan of one of the V1751 digitizer modules for testing purposes.








\end{document}